\documentclass[twocolumn]{article}
\usepackage{graphics}
\usepackage{graphpap}
\newcommand{\be}{\begin{equation}}
\newcommand{\ee}{\end{equation}}

\begin{document}
%
\twocolumn[\hsize\textwidth\columnwidth\hsize\csname @twocolumnfalse\endcsname
%
%
%

\title{The Ferromagnetic Kondo Model for
Manganites: \\ Phase Diagram and Charge Segregation Effects.}
\author{ E. Dagotto, S. Yunoki, A. L. Malvezzi, A. Moreo, and J. Hu \\
National High Magnetic Field Lab and Department of Physics,\\
Florida State University, Tallahassee, FL 32306  
\and
S. Capponi, and D. Poilblanc \\
Laboratoire de Physique Quantique 
Unit\'e Mixte de Recherche 5626, C.N.R.S., \\
Universit\'e Paul Sabatier, 31062 Toulouse, France
\and
N. Furukawa \\
Institute for Solid State Physics, \\ University of Tokyo,
Roppongi 7-22-1, Minato-ku, Tokyo 106, Japan }
\date{\today}
\maketitle

\begin{abstract}

The phase diagram of the ferromagnetic Kondo model for manganites is
investigated using computational techniques. In clusters of dimensions 1 and 2,
Monte Carlo simulations in the limit where the localized spins are
classical show a rich low temperature phase diagram with three dominant regions:
(i) a ferromagnetic phase, (ii) phase separation
between hole-poor antiferromagnetic and hole-rich ferromagnetic 
domains, and (iii) a phase with incommensurate spin correlations.
Possible experimental consequences of the regime of phase separation are
discussed.
Studies using the Lanczos
algorithm and the Density Matrix Renormalization Group method
 applied to chains with localized spin 1/2 
(with and without Coulombic
repulsion for the mobile electrons) and spin 3/2 degrees of freedom
give results in excellent
agreement with those in the spin localized classical limit.
The Dynamical Mean Field ($D=\infty$) approximation was also applied to
the same model. At large Hund coupling 
phase separation and ferromagnetism were identified,
again in good agreement with results in low dimensions.
In addition, a Monte Carlo study of spin correlations allowed us
to estimate the critical temperature for
ferromagnetism $T_c^{FM}$ in 3 dimensional clusters. 
It is concluded that $T_c^{FM}$ is compatible
with current experimental results. 

\end{abstract}
\vskip2pc]

%
%

\section{Introduction and Main Results}

Materials that present the phenomenon
of ``colossal'' magnetoresistance are currently under much experimental
investigation due to their potential technological applications.
Typical compounds that have this phenomenon are 
ferromagnetic (FM) metallic oxides of the form
 ${\rm R_{1-x} X_x Mn O_3}$ 
$({\rm where~ R= La,Pr,Nd;}$ ${\rm~X=Sr,}$ 
${\rm Ca,Ba,Pb})$~\cite{jin,tokura}. 
As an example, a decrease in resistivity of several
orders of magnitude has been reported 
in thin films of ${\rm Nd_{0.7} Sr_{0.3} Mn
O_3}$ at magnetic fields of $8$ Teslas~\cite{xiong}.
The relative changes in resistance
for the manganites can be as large as $\Delta R/R \sim 100,000 \%$, 
while in magnetic superlattices ${\rm Co/Cu/Co}$ the enhancement
is about $100\%$. 
This result suggests that manganites indeed have
technological potential since
large changes in resistance can be obtained at fixed temperature  upon
the application of magnetic fields, opening an alternative route for next
generation magnetic storage devices. 
However, since the development of La-manganite
sensors is still at a very early stage, 
a  more fundamental approach to the study of manganites
is appropriate and, thus, theoretical guidance is needed.
The existence of correlation effects in the fairly dramatic magnetic,
transport, and magneto-transport properties of doped La-manganites
reinforces this notion. 

The early theoretical studies of models for manganites concentrated their efforts on 
the existence of ferromagnetism. The so-called ``Double Exchange'' (DE)
model~\cite{zener,degennes} 
explained how carriers improve their kinetic energy by forcing the
localized spins to become ferromagnetically ordered (this phenomenon is
quite reminiscent of the Nagaoka phase  discussed in models for cuprates).
However, in spite of  this successful  explanation of the existence of
ferromagnetism at low temperature several features of the
experimental phase diagram of manganites remain unclear, and they are 
likely beyond the DE model. 
Actually, the phase diagram of ${\rm
La_{1-x} Ca_x Mn O_3}$ is very rich with not only ferromagnetic phases,
but also
regions with charge-ordering and antiferromagnetic correlations 
at ${\rm x > 0.5}$~\cite{phase}, and a poorly understood ``normal''
state  above the critical temperature for ferromagnetism, $T_c^{FM}$,
which has insulating characteristics at ${\rm x \sim 0.33}$. 
Finding an insulator above $T_c^{FM}$ is a surprising result since
it would have been more natural to have a standard metallic phase in
that regime
which could smoothly become a ferromagnetic metal as the
temperature is reduced. Some theories for manganites propose that the
insulating regime above $T_c^{FM}$ is caused by a strong correlation between electronic
and phononic degrees of freedom~\cite{millis}. Other proposals include the presence of
Berry phases in the DE model that may lead to electronic 
localization~\cite{muller}. On the other hand, the regime of charge
ordering has received little theoretical 
attention and its features remain mostly
unexplored. To complicate matters further, recent experiments testing
the dynamical response of manganites have reported anomalous results in
the ferromagnetic phase using
neutron scattering~\cite{lynn},
while in photoemission
experiments~\cite{dessau}  the  possible existence of a pseudogap
above the critical temperature was reported.

The  rich phase diagram of the
manganites described above, plus the several experimental indications of strong
correlations in the system, deserves a systematic theoretical study
using state-of-the-art computational techniques. These methods are
unbiased and can provide useful information on models for manganites
in a regime of couplings
that cannot be handled perturbatively or exactly. However, the large
number of degrees of freedom and associated couplings of a full
Hamiltonian model for manganites makes this approach quite cumbersome. 
In principle the two $e_g$ active orbitals per Mn-ion
 must be included, in
addition to the $t_{2g}$ electrons. Also phonons should be incorporated
to fully
describe these materials. However,
as a first step towards a theoretical understanding of the behavior
of models for manganites, in this paper it was decided to work only in the
electronic sector (i.e. leaving aside phonons) and with just one
orbital per site (i.e. keeping only one $e_g$ orbital of the
two available, which in practice amounts to assuming a static Jahn-Teller
distortion). In addition, the $t_{2g}$ degrees of freedom are here assumed
to be localized i.e. no mobility is given to these electrons. They
basically provide a spin background in which the $e_g$ electrons move,
with a Hund term that couples all active electrons per Mn-ion.
Under these assumptions a rich
phase diagram was obtained, as explained in detail
in the rest of the paper.
It is left for future work
the analysis of the influence of phonons and orbital degeneracy into the
fairly complicated phase diagrams reported here.

Some of the main results of the present effort
 are summarized in Figs.1 and 2 where the
phase diagrams of the ferromagnetic Kondo model, using classical spins to
represent the $t_{2g}$ spins, are presented for low dimensional systems.
These figures
 were reported previously in a short version of this paper~\cite{previous}
but here considerable
more details, as well as a large set of novel results, are provided.
Three dominant regimes have been
identified: (1) a ferromagnetic region 
in excellent agreement with the DE mechanism, (2)
phase separation between hole-rich ferromagnetic
and hole-poor antiferromagnetic regions, and (3) an intermediate
phase with short-range incommensurate correlations.
The regime of phase separation was previously
 conjectured to exist in this model
from  studies of  Hamiltonians for manganites at large Hund coupling
in one dimension~\cite{jose}.
It is important to remark that phase separation is currently
widely discussed in the context of high temperature superconductors
since models for the cuprates, such as the $t-J$ model in two dimensions,
 present densities that cannot be uniformly
stabilized in a given volume by suitably selecting a chemical 
potential~\cite{tj1,review,tj3}. After the introduction of
$1/r$ Coulombic interactions the resulting
hole-rich regions become unstable against the formation of
``stripe'' configurations~\cite{tj2,rice}. If the tendencies
towards  phase separation
reported in this paper are realized in the manganites, a similar 
phenomenon may likely occur
i.e. neutron scattering experiments could reveal
evidence of stripe configurations in compounds such as 
${\rm La_{1-x} Ca_x Mn O_3}$ as it occurs in the cuprates~\cite{tranquada}.
The pseudogap features found in
photoemission~\cite{dessau} could also be related to this phenomenon.
\begin{figure}[h]
{\rotatebox{0}{
{\resizebox{8.5cm}{!}{\includegraphics{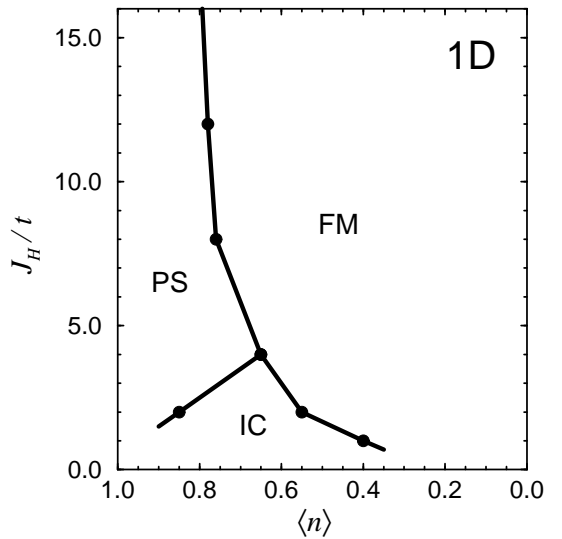}}}
}}
\vspace*{-3cm}
\caption{
Phase diagram of the ferromagnetic Kondo model
in one dimension obtained with Monte
Carlo techniques in the limit where the localized spins are classical.
PS,~FM, and IC denote regions with phase separation,
ferromagnetism, and incommensurate correlations. For details see the
text.
}
\end{figure}

The bulk of the paper is devoted to
the discussion of
the numerical evidence that provides support to these proposed phase
diagrams. Results were not only obtained in dimensions 1 (Secs.III and
IV)
and 2 (Sec.V), but also
in 3 (Sec.VII) and $\infty$ (Sec.VI) 
using a variety of numerical techniques. Both classical and quantum
mechanical $t_{2g}$ degrees of freedom were analyzed on chains.
Experimental consequences of our results are discussed, specially those
related with the existence of phase separation (Sec.VIII).
There is a clear underlying 
qualitative universality between results obtained on lattices with
different coordination number 
and using different algorithms. This universality lead us to believe
that the results reported here contain at least the 
dominant  main features
of the phase diagram corresponding to realistic electronic models for manganites that have
been widely discussed before in the literature, but which were not
systematically studied using computational methods.
\begin{figure}
{\rotatebox{0}{
{\resizebox{8.25cm}{!}{\includegraphics{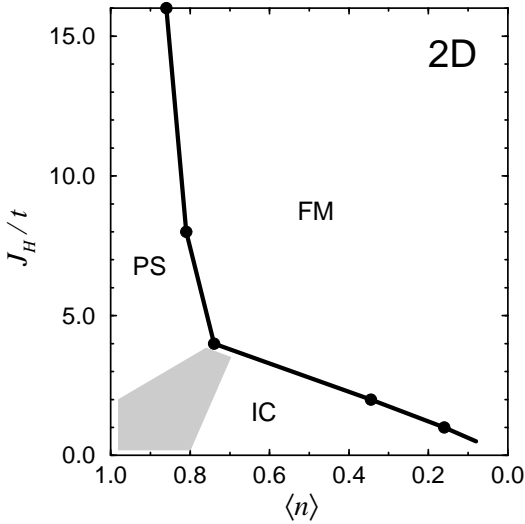}}}
}}
\vspace*{-3cm}
\caption{
Same as Fig.1 but for the case of two dimensions. The boundary
between the PS and IC regimes was difficult to obtain numerically and,
thus, it is not sharply defined in the figure (gray region).
}
\end{figure}

\section{Model, Symmetries and Algorithm}

The ferromagnetic Kondo Hamiltonian~\cite{zener,furukawa} studied in
this paper is defined as 
$$
H = -t \sum_{\langle {\bf i j} \rangle \sigma} (c^\dagger_{{\bf i}\sigma} c_{{\bf
j}\sigma} + h.c.) - J_H \sum_{{\bf i}\alpha \beta}
{ { c^{\dagger}_{{\bf i}\alpha} {\bf \sigma}_{\alpha \beta} c_{{\bf
i}\beta} }
\cdot{{\bf S}_{\bf i}}},
\eqno(1)
$$
\noindent where $c_{{\bf i}\sigma}$ are destruction
operators for electrons at site ${\bf i}$
with spin $\sigma$,
and ${\bf S}_{\bf i}$ is the total 
spin of the $t_{2g}$ electrons, assumed 
localized.  
The first term is the electron transfer between nearest-neighbor
Mn-ions,
$J_H>0$ is the ferromagnetic Hund coupling, the number of sites is $L$,
and the rest of the notation is standard. The boundary conditions used
in the present study are important for some results, and they will be discussed
later in the text.
The electronic
 density of $e_g$ electrons, denoted by $\langle n \rangle$, is adjusted using
a chemical potential $\mu$.
In several of the results shown below
 the spin ${\bf S}_{\bf i}$  will be considered classical
(with $|{\bf S}_{\bf i}| = 1$). In 1D both classical and
quantum
mechanical $t_{2g}$ spins will be studied, the latter having a realistic
spin 3/2 but also considering spin 1/2 for comparison.
Phenomenologically $J_{H} \gg t$, 
but here $J_H/t$ was
considered an arbitrary parameter, i.e. both large and small values
for $J_H/t$ were studied. Below some calculations were also carried out
including a large on-site Coulombic repulsion among the mobile electrons.

For a one-dimensional chain (open ends), or for a one-dimensional ring with
$L$ even and periodic or antiperiodic boundary conditions (PBC and
APBC, respectively)
the model is particle-hole symmetric with respect to $\langle n \rangle
=1$ by simply transforming $c^\dagger_{{\bf i}\uparrow}
\rightarrow (-1)^{\bf i}  c_{{\bf i}\downarrow}$ 
and $c^\dagger_{{\bf i}\downarrow}
\rightarrow -(-1)^{\bf i}  c_{{\bf i}\uparrow}$
for the mobile electrons. In this case
the density is transformed as $\langle n \rangle \rightarrow 2 - \langle
n \rangle$. Similar transformations can be deduced for clusters of dimension
larger than one. Then, here
it is sufficient to study densities $\langle n \rangle \leq 1$.

The FM Kondo model Eq.(1) with classical spins can be substantially
simplified if the limit $J_H \rightarrow \infty$ is also considered.
In this situation at every site
only the spin component of the mobile electrons
which is parallel to the classical spin is a relevant 
degree of freedom. The best way to make this reduction in the Hilbert
space is by rotating the $c_{{\bf i}\sigma}$ operators into new
operators $d_{{\bf i}\alpha}$ using a $2 \times 2$ rotation matrix
such that the transformed
spinors  point in the direction of the classical spin.
Explicitly, the actual transformation is:
$$
\hspace{1.52cm} c_{{\bf i}\uparrow}=\cos(\theta_{\bf i} /2)~d_{{\bf i} \uparrow}
-\sin(\theta_{\bf i} /2)e^{-i \phi_{\bf i}}~d_{{\bf i}\downarrow},
\eqno(2.1) 
$$
$$
c_{{\bf i}\downarrow}=\sin(\theta_{\bf i}/2)e^{i \phi_{\bf i}}
~d_{{\bf i}\uparrow}+\cos(\theta_{\bf i}/2)~d_{{\bf i}\downarrow}.
\eqno(2.2)
$$
%
\noindent The angles $\theta_{\bf i}$ and $\phi_{\bf i}$ define
the direction of the classical spin at site ${\bf i}$.
After this transformation the new Hamiltonian becomes
$$
H_{J_H = \infty} = - \sum_{\langle {\bf m,n} \rangle} ( t_{\bf m,n}
d^\dagger_{{\bf m}\uparrow} d_{{\bf n}\uparrow} + h.c.),
\eqno(3)
$$
\noindent where the down component of the new operators has been
discarded since the
$J_{H} \rightarrow \infty$ limit is considered. The effective hopping is
\setcounter{equation}{3}
\begin{eqnarray}
t_{\bf m,n} & = & t\,[\cos(\theta_{\bf m}/2)
\cos(\theta_{\bf n}/2)  \nonumber   \\
            &   & +\sin(\theta_{\bf m}/2)\sin(\theta_{\bf n}/2)
 e^{-i(\phi_{\bf m}-\phi_{\bf n})} ] ,
\end{eqnarray}
\noindent i.e. it is complex and dependent on the 
direction of the classical spins at sites ${\bf m,n}$~\cite{muller}.
In the limit $J_H = \infty$ the band for the transformed spinors
 with up spin has itself a
particle-hole symmetry, which exists for any arbitrary
configuration of classical spins. This can be shown by transforming
$d^\dagger_{{\bf i}\uparrow} \rightarrow (-1)^{\bf i}  
d_{{\bf i}\uparrow}$, and noticing that after this transformation 
the resulting Hamiltonian matrix
is simply the conjugate of the original. Then,
the eigenvalues are unchanged, but $\langle 
d^\dagger_{{\bf i}\uparrow}
d_{{\bf i}\uparrow} \rangle \rightarrow 1 - \langle d^\dagger_{{\bf i}\uparrow}
d_{{\bf i}\uparrow} \rangle$.

The partition function of the FM Kondo model with classical spins
can be written as 
\setcounter{equation}{4}
\begin{equation}
Z =  \prod^{L}_{\bf i} \left(\,\int^{\pi}_0 d\theta_{\bf i} \sin\theta_{\bf i}
\int^{2 \pi}_0 d\phi_{\bf i}\,\right)\,\mathrm{tr}_{c}(e^{-\beta H} ).
\end{equation}
For a fixed configuration of angles
$\{ \theta_{\bf i}, \phi_{\bf i} \}$ 
the Hamiltonian amounts to
non-interacting electrons moving in an external field. This problem
can be diagonalized exactly since the Hamiltonian is quadratic in the
fermionic variables. This diagonalization is performed simply by calling 
a library subroutine in a computer program. 
If the $2L$ eigenvalues are denoted by
$\epsilon_{\lambda}$ the resulting partition function becomes
\begin{equation}
Z =  \prod^{L}_{\bf i} \left(\,\int^{\pi}_0 d\theta_{\bf i} \sin\theta_{\bf i}
\int^{2 \pi}_0 d\phi_{\bf i}\,\right)\,\prod^{2L}_{\lambda = 1} 
(1 + e^{-\beta
\epsilon_{\lambda}} ).
\end{equation}
\noindent The integrand in Eq.(6) is positive and, thus, a Monte Carlo
simulation can be performed on the classical spin angles without
``sign'' problems.
This was the procedure followed in our study below. The number of sweeps
through the lattice needed to obtain good statistics
 varied widely depending on the temperature, densities, and couplings.
In some cases, such as in the vicinity of phase separation regimes, up to
$10^6$ sweeps were needed to collect good statistics.
Measurements of equal-time
spin and charge correlations for the mobile electrons
 were performed by transforming the
operators involved using the basis that diagonalizes the Hamiltonian
for a fixed configuration of classical spins. Dynamical studies of the optical
conductivity $\sigma(\omega)$ and the one particle spectral function
$A({\bf p},\omega)$ could also be performed following a similar procedure,
 but their detailed study is postponed for a future publication.
The analysis of the FM Kondo model in the case of quantum mechanical
$t_{2g}$ degrees of freedom was performed at zero temperature using
standard Lanczos~\cite{review,didier} 
and Density Matrix Renormalization Group (DMRG)~\cite{white} techniques.
The study at infinite dimension was carried out with the Dynamical Mean-Field
approach~\cite{furukawa}.

\section{Results in D=1 with Classical Localized Spins}

In this section the computational results that led us to propose Fig.1 as the
phase diagram of the FM Kondo model in one dimension using classical spins are presented.
Results for quantum mechanical $t_{2g}$ spins are also provided.

\subsection{Ferromagnetism}

The boundary of the ferromagnetic region in Fig.1 was
obtained by studying the spin-spin correlations (between the classical
spins) defined in momentum space as
$S({ q}) = 
(1/L) \sum_{ j,m} e^{i {{ (j-m)}\cdot{ q}}}
\langle {{{\bf S}_{ j}}\cdot{{\bf S}_{ m}} } 
   \rangle$, using  a standard notation, at the particular value of 
zero momentum. $L$ is the number of sites.
The analysis was performed for couplings $J_H/t = 1,2,3,4,8,12$ and $18$.
For the last four (strong) couplings and in the region of densities that
were found to be stable (see next section 
for a discussion) the real space spin-spin
correlations $\langle {{{\bf S}_{ j}}\cdot{{\bf S}_{ m}} }    \rangle$
 have a robust tail that extends to the
largest distances available on the clusters studied here. This is
obviously correlated with the presence of a large peak in $S(q)$
at zero momentum. Fig.3
shows representative results at $J_H/t=12$ using open boundary
conditions (OBC) and $L=24$. 
A robust ferromagnetic
correlation is clearly observed even at large distances. The strength
of the tail
decreases as $\langle n \rangle$ increases in the region $\langle n
\rangle \geq 0.5$,
and it tends to vanish at a density
$\sim 0.78$ which corresponds to the boundary of the 
ferromagnetic region in Fig.1. A similar behavior is observed
for smaller values of the coupling.
It is interesting to note that the maximum strength of the
spin-spin correlation tail appears at $\langle n \rangle \approx 0.50$.
This result is compatible with the behavior expected at $J_H = \infty$ where 
large and small densities are exactly related by
symmetry (as discussed in Sec.II)
and, thus, the ferromagnetic correlations should peak at exactly
$\langle n \rangle = 0.50$. 
Working at $J_H/t \leq 8$ this qualitative behavior is washed out and in
this regime
the correlations at densities $\langle n \rangle \leq 0.50$ are very similar.

Special care must be taken with
the boundary conditions (BC).
Closed shell BC or open BC are needed
to stabilize a ferromagnetic spin arrangement. If other BC are used the
spin correlations  at short distances
are still strongly FM (if working at couplings where ferromagnetism is
dominant), but not at large distances where they become negative
since a kink appears separating two regions of opposite total spin, with
each region having all spins aligned in the same direction.
This
well-known effect was observed before in a similar context~\cite{jose,zang}
and it does not present a problem in the analysis shown below.
Actually results using
other boundary conditions and lattice lengths up to 60 sites
are  compatible with
the data of Fig.3. 
In particular it can be shown that the ferromagnetic correlations 
persist when the lattice size is increased. Fig.4a shows $S(q=0)$ versus
temperature for several lattice sizes. At small temperature the
sum over sites of the correlations grow with $L$ showing that the FM correlation
length is larger than the size of the chains used here.
However, in 1D the Mermin-Wagner theorem forbids
the existence of a finite critical temperature and, thus, eventually
$S(q=0)$ must tend to saturate at any fixed finite temperature
$T$ as $L$ increases, as it already
occurs for $T \geq 0.02$ in Fig.4a.
In spite of this subtle detail, 
using closed-shell boundary conditions the numerical data 
presented in this subsection supports 
the presence of
strong ferromagnetic correlations in the
one dimensional FM Kondo model, and it is reasonable to expect
 that at zero
temperature the model in the bulk
 will develop a finite magnetization in the ground state.
\begin{figure}[h]
{\rotatebox{0}{
{\resizebox{7.25cm}{!}{\includegraphics{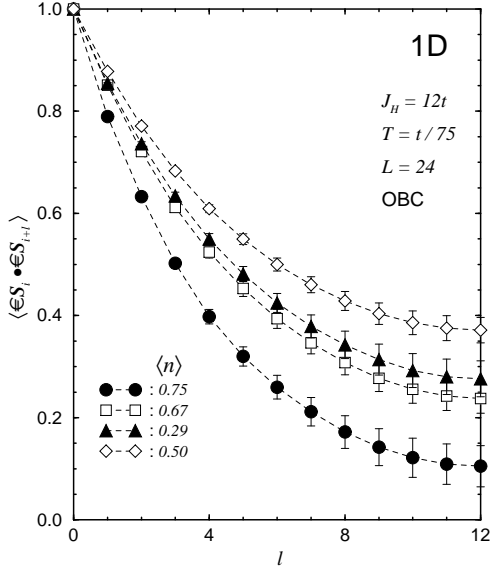}}}
}}
\vspace*{-1.5cm}
\caption{
Spin-spin correlations among the localized spins (assumed
classical) obtained with the Monte Carlo method
 working on a 1D system, and at the coupling, temperature, densities, and
lattice size shown. Open boundary conditions were used.
Clear strong ferromagnetic correlations are observed.
}
\end{figure}
Fig.4b contains the spin-spin correlations vs distance parametric with
temperature. They show that even at relatively high temperatures, the
correlations at $short$ 
distances are clearly ferromagnetic, an effect
that
should influence on transport properties since carriers will react
mainly to the local environment in which they are immersed.
Fig.5 contains information that illustrates the formation of
ferromagnetic spin
polarons at relatively high temperatures. Fig.5a
shows the spin correlations at low electronic density. At
$T=t/10$ and distance 1, the correlation is only about 15\%
of the maximum. However,
if this correlation is measured in the immediate vicinity of a
carrier using $\langle n^e_i { {{\bf S}_{i}}\cdot{{\bf S}_{i+l}}} \rangle$,
where $n^e_i = \sum_{\sigma} n_{i \sigma} (1 - n_{i -\sigma})$ and
$n_{i \sigma}$ is the number operator for $e_g$-electrons at site $i$
and with spin projection $\sigma$,
then the correlation is enhanced to 40\%. Thus, it is clear that FM
correlations develop in the vicinity of the carrier~\cite{horsch}. 
This spin polaron
apparently has a size of 3 to 4 lattice spacings in our studies.
\begin{figure}
{\rotatebox{0}{
{\resizebox{7.75cm}{!}{\includegraphics{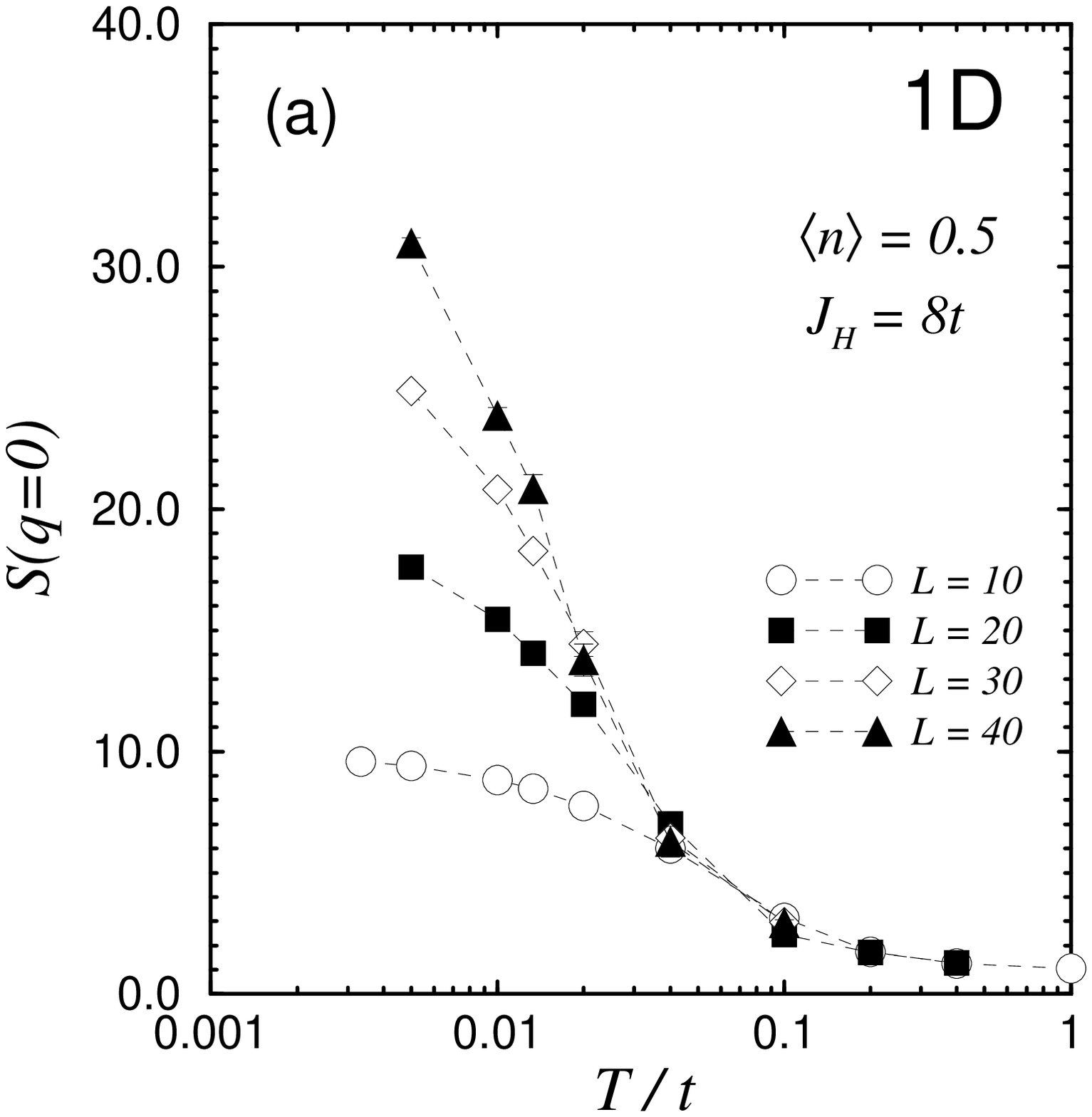}}}
}}
{\rotatebox{0}{
{\resizebox{7.75cm}{!}{\includegraphics{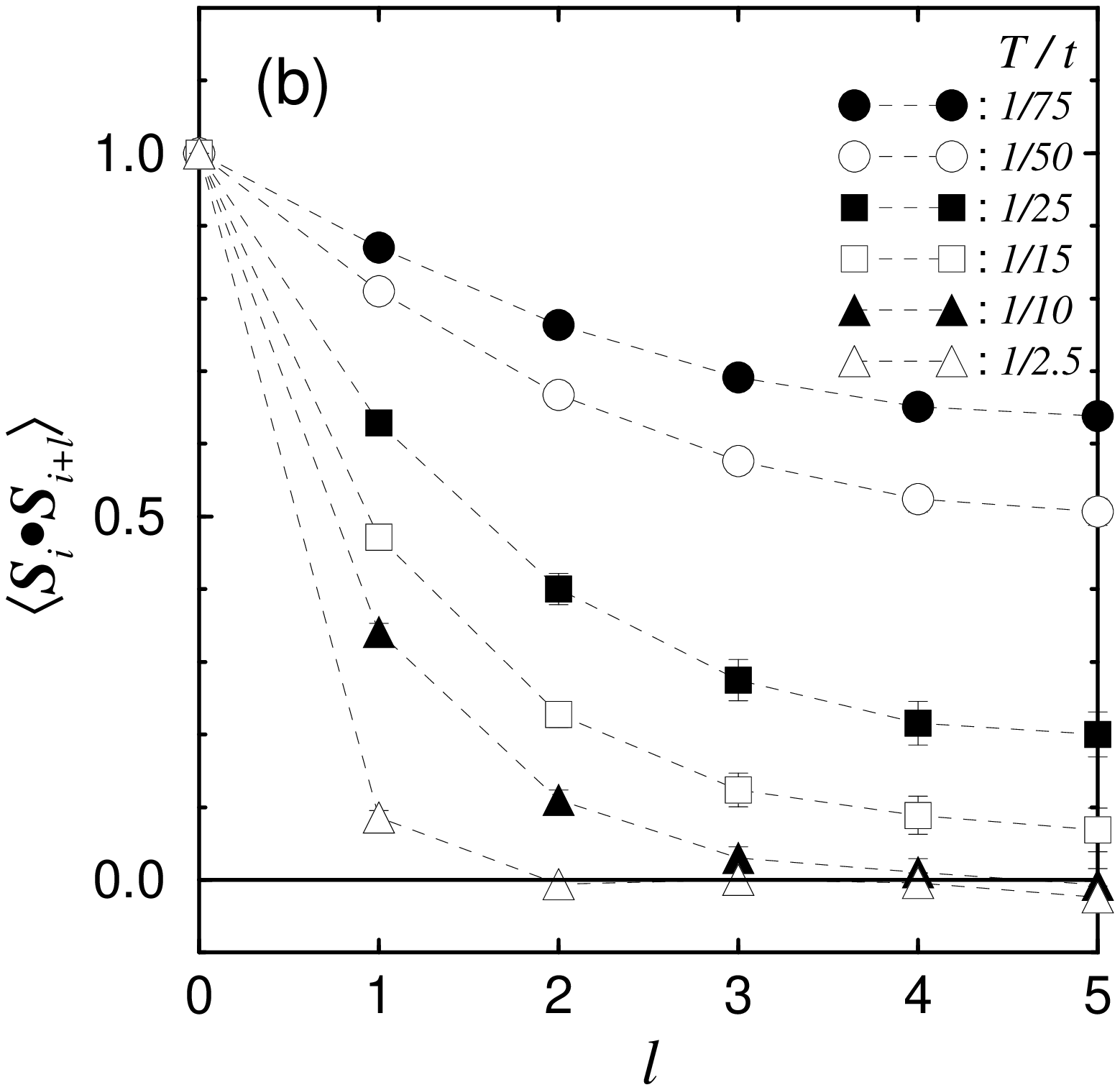}}}
}}
\vspace*{-2.cm}
\caption{
(a) Temperature dependence of the spin correlations in the one
dimensional FM Kondo model using classical spins and the Monte Carlo
method. Shown is the
zero momentum spin structure factor $S(q)$ (for the classical spins)
at the density, coupling, and lattice sizes indicated. Closed shell boundary
conditions were used (periodic boundary conditions (PBC)
for $L=10$ and 30, and antiperiodic boundary conditions (APBC) for
$L=20$ and 40); (b) Spin-spin correlations at density $\langle n \rangle
= 0.7$, using a 10 site chain, PBC, and $J_H/t = 8.0$. The temperatures
are indicated.
}
\end{figure}
\begin{figure}
{\rotatebox{270}{
{\resizebox{6.5cm}{!}{\includegraphics{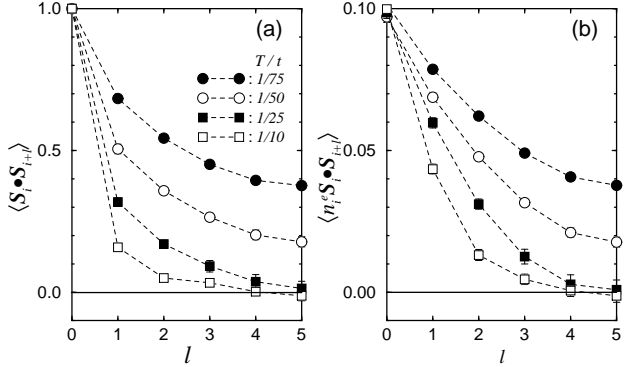}}}
}}
\caption{
(a) Spin-spin correlations at $\langle n \rangle = 0.1$ using
a chain of 10 sites with PBC. The Hund coupling is $J_H/t = 8.0$; (b)
Similar as (a) but measuring the spin correlations near a conduction
electron using $\langle n^e_i { {{\bf S}_{i}}\cdot{{\bf S}_{i+l}}}
\rangle$,
where $n^e_i$ is defined in the text.
}
\end{figure}

\subsection{Phase Separation}

One of the main purposes of this paper
is to report the existence of phase separation in the FM Kondo model.
To study this phenomenon in the grand-canonical  ensemble, where the
Monte Carlo simulations are performed, it is
convenient to analyze $\langle n \rangle$ versus $\mu$. If
$\langle n \rangle(\mu)$ is discontinuous then there are densities that can not
be estabilized, regardless of the value of $\mu$.
The results shown in Fig.6 obtained
at $J_H/t=12$ clearly show that indeed phase separation
occurs in the FM Kondo model.
The discontinuity is between
a density corresponding to the antiferromagnetic
regime $\langle n \rangle = 1.0$ and $\sim 0.77$ where ferromagnetic 
correlations start developing, as shown before in Fig.3. 
In the case of the canonical ensemble 
these results can be rephrased as follows:
if the system is initially setup with a
density in the forbidden band, it will spontaneously separate into two
regions having (i) antiferromagnetic (AF)
 correlations and no holes, and 
(ii) FM correlations and most of the holes.

The existence of phase separation can also be deduced from
the actual Monte Carlo runs since they require 
large amounts of CPU time for convergence
in the vicinity of the critical chemical potential $\mu_c$. 
The reason is that
in this regime there are two states
in strong competition. Qualitatively this effect can be visualized
analyzing $\langle n \rangle$ as a function of Monte Carlo time. Fig.7 shows
such a time evolution when the chemical potential is fine tuned to its
critical value $\mu_c \sim -6.69812t$ at $J_H/t=8$. Wild fluctuations
in $\langle n \rangle$ are observed with frequent tunneling events 
covering a large range of densities. A change in $\mu$ as small as $0.001$ or
even smaller reduces drastically the frequency of the
 tunneling events, and makes the results
\begin{figure}[h]
{\rotatebox{270}{
{\resizebox{7.25cm}{!}{\includegraphics{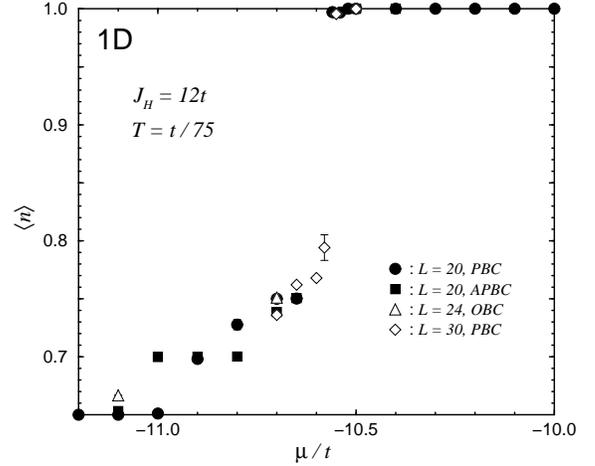}}}
}}
\caption{
Electronic density $\langle n \rangle$ vs the chemical potential
$\mu$ obtained with the Monte Carlo technique applied to the one
dimensional FM Kondo model with classical spins. Coupling,
temperature, lattice sizes, and boundary conditions
are indicated. The discontinuity suggests that some
densities are unstable, signaling the presence of phase separation.
}
\end{figure}
\begin{figure}[!h]
\vspace*{-.5cm}
{\rotatebox{270}{
{\resizebox{7.75cm}{!}{\includegraphics{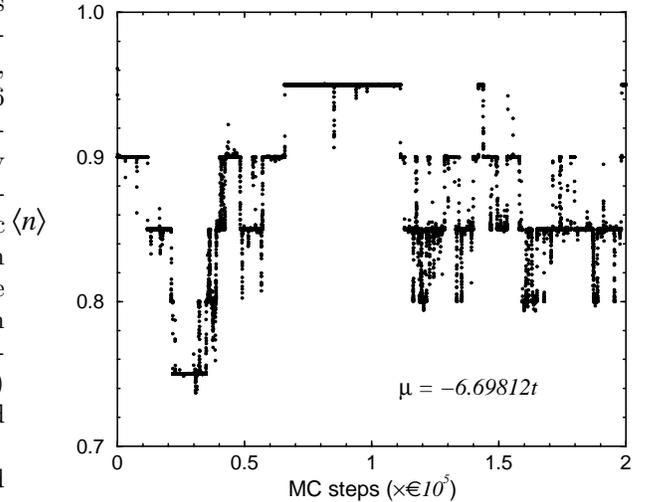}}}
}}
\caption{
Monte Carlo time evolution of the density $\langle n \rangle$ at the
particular value of $\mu$ where the discontinuity takes place working
at $J_H/t=8$, $T=t/75$, using $L=20$ sites, and PBC.
Frequent tunneling events are observed showing the competition between
two states as in a first order phase transition.
}
\end{figure}
more stable although certainly strong
fluctuations remain in a finite window near $\mu_c$. 
Fig.8 illustrates this effect showing a histogram
that counts the number of times that a density in a given
window of density is reached in the simulation. 
As $\mu$
crosses its critical value
the histograms change rapidly 
 from having a large peak close to $\langle n
\rangle \sim 1$ to a large peak at $\langle n \rangle \sim 0.75$. At
densities that are stable away from the phase separation region these
histograms present just one robust peak.

The qualitative behavior exemplified in Fig.6 was also observed at
other values of $J_H/t$. For instance, Fig.9a contains resuls for
$J_H/t=4$ which are very similar to those found
at a larger coupling. In a weaker coupling regime the discontinuity is
reduced and now the competition is between the AF state and a state with
incommensurate correlations rather than ferromagnetism. 
As example, Fig.9b contains
results for $J_H/t=2$. The discontinuity is located near $\mu_c \sim
-1.35t$
and the two states competing in this regime actually have very
similar properties. Analyzing, for instance, just 
 $S(q)$ would have not
provided an indication of a sharp discontinuity in the density, 
unlike the case of a
large Hund coupling where the peak in the spin structure factor jumps
rapidly from $q=\pi$ to $0$ at the critical chemical potential.
\begin{figure}[!h]
{\rotatebox{0}{
{\resizebox{7.75cm}{!}{\includegraphics{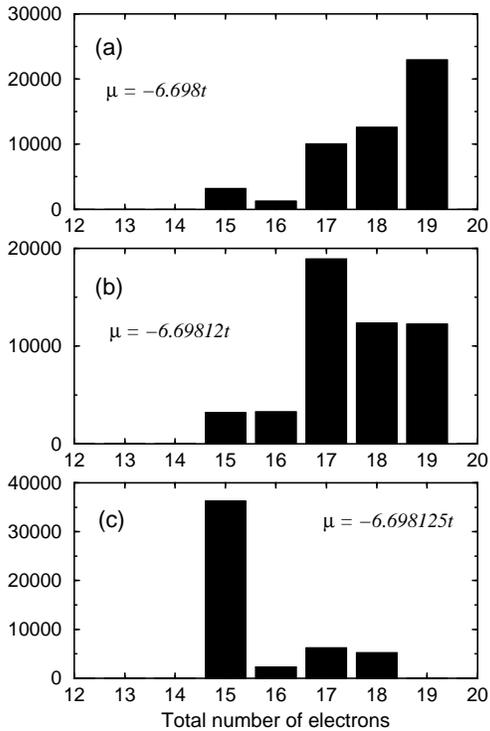}}}
}}
\caption{
Hystogram for the total number of electrons obtained at $J_H/t =
8$, $T=t/75$, using periodic boundary conditions on a 20 sites chain.
The chemical potentials are shown. (a) corresponds to a case where the
average density is close to $\langle n \rangle =1$, (b) is at the
critical chemical potential, and at (c)
the electronic density is close to 0.77.
}
\end{figure}
%
%
\begin{figure}[!h]
{\rotatebox{270}{
{\resizebox{6.5cm}{!}{\includegraphics{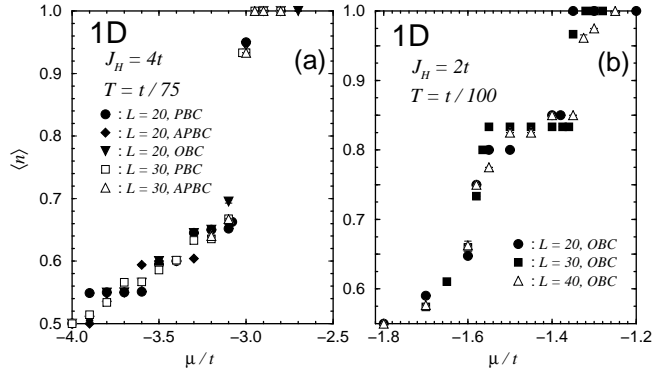}}}
}}
\caption{
Electronic density vs $\mu$ for the FM Kondo model with classical spins
in one dimension using Monte Carlo methods. A clear discontinuity
signals the existence of phase separation. (a) corresponds to $J_H/t =
4$, while (b) is for $J_H/t =2$.
Temperatures, chain sizes, and boundary conditions are indicated.
The apparent second discontinuity located at
$\mu/t \sim -1.55$ in (b) is expected to be
just a rapid crossover.
}
\end{figure}

\subsection{Incommensurate Correlations}

The tendency to develop a spin
pattern with incommensurate characteristics can be easily studied in
our calculations observing the behavior of $S(q)$ as couplings and
densities are varied. While at large $J_H/t$ the spin structure factor
is peaked only at the momenta compatible with ferro or antiferromagnetic
order, a different result is obtained as $J_H/t$ is reduced. Fig.10
shows $S(q)$ at $J_H/t = 1.0$ for a variety of densities. 
The AF peak close to $\langle n \rangle = 1$ smoothly evolves
into a substantially weaker peak which moves away from $q=\pi$  in
the range $0.5 \leq \langle n \rangle \leq 1.0$. 
The peak position is close to $2 k_F = \langle n \rangle$, and since the
spin-spin correlations for the mobile electrons show a similar behavior
this is compatible with 
Luttinger liquids predictions~\cite{voit}.
In this density
range, and for the lattice sizes and temperatures used here,
the peak in $S(q)$ is broad and, thus, it cannot be taken as an
indication of long-range incommensurate (IC) correlations 
but rather of the presence of IC
spin arrangements  at short distances. In the region
of low densities
$\langle n \rangle \leq 0.4$, $S(q)$ is now peaked at zero momentum 
which is compatible with the presence of robust ferromagnetic
spin-spin correlations at the largest distances available in the studied
clusters. The transition from one regime to
the other in the intermediate small window
$0.4 \leq \langle n \rangle \leq 0.5$ is very fast and
was not studied in detail here,
but for a second order phase transition it is expected to be continuous.
In Fig.11 results for a slightly larger coupling $J_H/t=2$ are shown.
Here the pattern is more complicated since apparently the AF
peak does not evolve smoothly into the peak at $q \sim \pi/2$ observed
at $\langle n \rangle = 0.733$, suggesting an interesting interplay
between spin and charge.
\begin{figure}[!h]
{\rotatebox{270}{
{\resizebox{7.75cm}{!}{\includegraphics{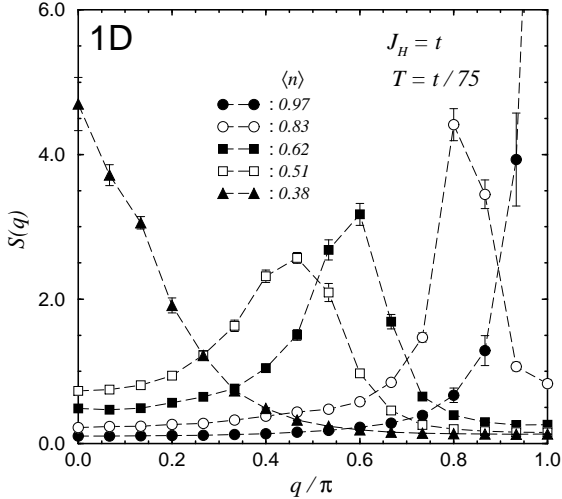}}}
}}
\caption{
$S(q)$ (Fourier transform of the spin-spin correlations between classical
spins) at $J_H = t$ and $T=t/75$, on a chain with 30 sites and PBC.
The densities are indicated.
The positions of the peaks indicate the tendency to have
incommensurate correlations in the FM Kondo model.
}
\end{figure}
\begin{figure}[!h]
{\rotatebox{270}{
{\resizebox{7.75cm}{!}{\includegraphics{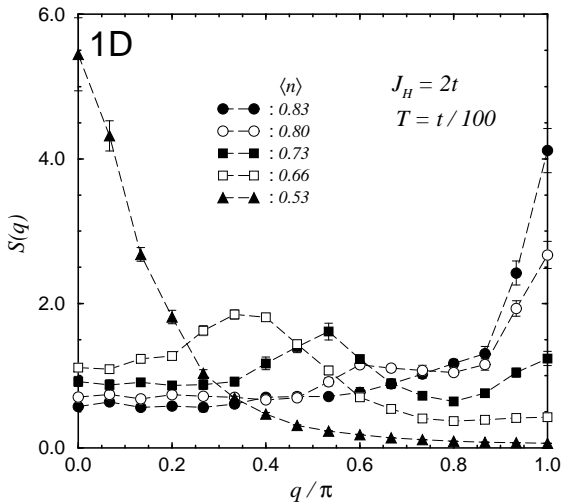}}}
}}
\caption{
Same as Fig.10, but working at $J_H = 2t$ and $T=t/100$, with 30
sites and open boundary conditions (OBC).
}
\end{figure}

Note that the presence of IC correlations in models for manganites
 was predicted theoretically using
a Hartree-Fock approximation~\cite{inoue}. Our results are compatible
with these predictions although, once again, it is not clear if the IC pattern
corresponds to long-range order or simply short distance correlations.
More work is needed to clarify these issues. Nevertheless, the present
effort is enough to show that the tendency to form IC spin patterns
exists in the FM Kondo model at small $J_H/t$.

\subsection{Influence of a Direct Coupling Among the Localized Spins}

The results of Sec.III.B show the presence of phase separation 
near half-filling, but not in the opposite extreme of low
conduction electron density. This is a consequence of the absence of a
direct coupling among the localized spins in Eq.(1). This
coupling may be caused by a small hybridization between
the $t_{2g}$ electrons. If a Heisenberg term
$J' \sum_{\langle i j \rangle} {{{\bf S}_{ i}}\cdot{{\bf S}_{ i}}}$
coupling the classical spins
is added, then at $\langle n \rangle = 0$ an antiferromagnetic state
is recovered similarly as at $\langle n \rangle = 1$.
\begin{figure}[!h]
{\rotatebox{270}{
{\resizebox{6.8cm}{!}{\includegraphics{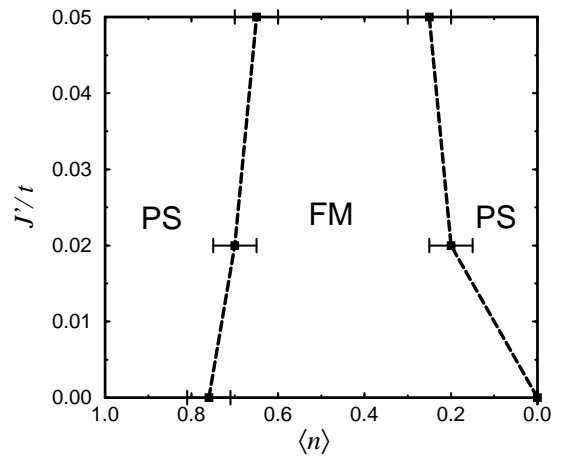}}}
}}
\caption{
Phase diagram in the plane $J'/t$-$\langle n \rangle$
obtained studying the density vs $\mu$ calculated using the Monte Carlo
technique with classical localized spins. The coupling is fixed to
$J_H/t =8$ and the temperature is low. Note that the regime of
low density and $J'/t = 0$ is somewhat difficult to study due to
the influence of van Hove singularities which produce a rapid change of
the density with $\mu$.
}
\end{figure}
This term was already considered in the study of the strong coupling
version of the Kondo model~\cite{jose}. Although a detailed study of
the influence of $J'$ on the phase diagram Fig.1 is postponed for a future
publication~\cite{yunoki2},
here the effects of this new coupling into the existence of phase
separation are reported. Following the same procedure described in
Sec.III.B
to obtain unstable densities, the phase diagram shown in Fig.12 was
found. Note that at $J' \neq 0$ phase separation occurs at large $and$
small electronic densities. In the latter the separation is
 between  electron-rich
ferromagnetic and electron-poor antiferromagnetic regions. Then, the
experimental search for phase separation discussed below in Sec.VIII
should be carried out at both large and small hole densities. More details
will be given elsewhere~\cite{yunoki2}.

\section{Results in D=1 with Quantum Localized Spins}

\subsection{Quantum Mechanical $t_{2g}$ S=3/2 spins}

The use of classical spins to represent the $t_{2g}$ degrees of freedom
is an approximation which has been used since the early days of the
study of manganites~\cite{zener,degennes}. 
While such an approach seems reasonable it would be
desirable to have some numerical evidence supporting the idea
 that using 
spin operators of value 3/2 (denoted by
${\bf {\hat S}}_{\bf i}$) the results are
similar as those obtained with classical spins. Although numerical unbiased
calculations with spins 3/2 are difficult in dimensions larger than 1,
at least 
this issue can be addressed numerically with 1D chains
using the Lanczos and Density Matrix Renormalization Group techniques.
The Hamiltonian is the same as in Eq.(1)
but now with quantum mechanical degrees of freedom normalized to 1 (i.e.
replacing
${\bf S}_{\bf i}$ by  ${\bf {\hat S}}_{\bf i}/(3/2)$) to simplify the 
comparison of results against those obtained using classical spins.
Calculating the ground state energy in subspaces with a fixed
total spin in the $z$-direction, it is possible to study the
tendency to have a ferromagnetic state in the model using the 
Lanczos technique.
The results indicate that there is a robust region of fully
saturated
ferromagnetism, as indicated in Fig.13. The actual
boundary of this region agrees accurately with
the results obtained using classical spins shown in Fig.1. 
This reinforces the belief that
S=3/2 and classical spins produce similar results, at least regarding
ferromagnetism.
Finite size effects are apparently small for the chains accessible with
the DMRG method, as exemplified in 
Fig.14a where the ground state energy for
$J_H/t=6.0$ using a variety of chain lengths is presented. 
\begin{figure}
{\rotatebox{0}{
{\resizebox{7.75cm}{!}{\includegraphics{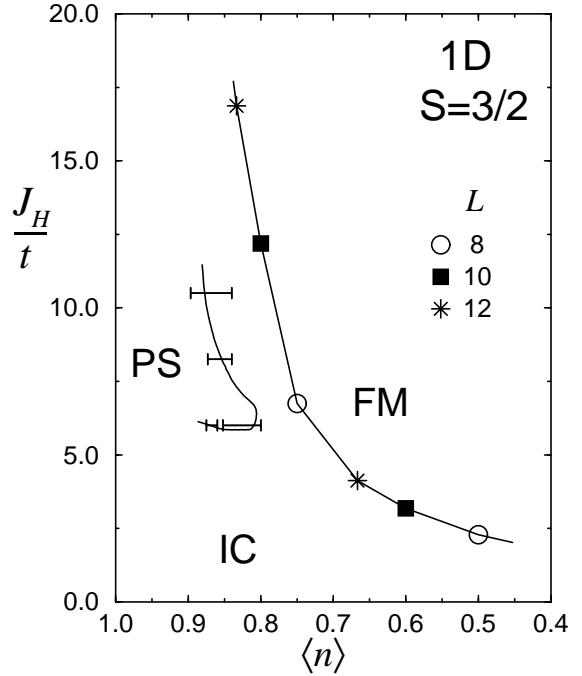}}}
}}
\vspace*{-0.5cm}
\caption{
Phase diagram of the FM Kondo model with $S=3/2$ localized $t_{2g}$
spins obtained with the
DMRG and Lanczos techniques
applied to finite chains as indicated. The notation is
as in Figs.1 and 2, i.e. ferromagnetism, incommensurability, and phase
separation appear both for classical and quantum mechanical localized spins.
}
\end{figure}
\begin{figure}
{\rotatebox{270}{
{\resizebox{5.75cm}{!}{\includegraphics{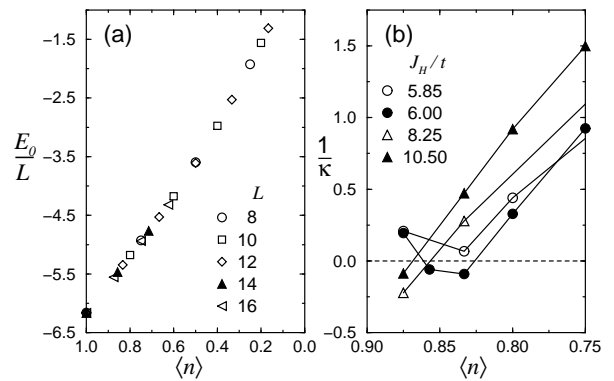}}}
}}
\caption{
(a) Ground state energy per site vs density at $J_H/t=6.0$ for
the $S=3/2$ FM Kondo model in 1D using DMRG techniques keeping
$m=48$ states. Results for
a variety of chain lengths are shown; (b) Inverse compressibility vs
density for the same model and chain lengths as in (a), calculated using
several couplings. A negative $1/\kappa$ signals an unstable density.
}
\end{figure}

To study other important features of the phase diagram, such as phase
separation, 
the compressibility is needed. This quantity is defined as
$$
\kappa^{-1} = {{N_e^2}\over{L}} {{E(N_e+2,L) + E(N_e-2,L) - 2E(N_e,L)}\over{4}},
\eqno(7)
$$
\noindent where $E(N_e,L)$ is the ground state energy corresponding to a
chain with $L$ sites and $N_e$ electrons. If $\kappa^{-1}$ becomes
negative at some fixed density (remember that 
here the numerical study is in the
canonical ensemble), the system is unstable and phase separation occurs.
The DMRG results for the compressibility  in Fig.14b for several
couplings led us to
conclude that the spin 3/2 model also has phase separation in the ground
state near half-filling, similarly as in the case of classical spins.
Indeed Fig.14b shows that
$1/\kappa$ becomes negative for densities
in the vicinity of $\langle n
\rangle \sim 0.85$ and larger,
 for the range of couplings shown. Thus, at least
qualitatively there are tendencies to phase separate in the same
region suggested by the simulations using classical localized spins.

However, a difference exists between results obtained with
classical and quantum mechanical
$t_{2g}$ degrees of freedom:
apparently there is a finite window in density between the phase separated
 and FM regimes in the phase diagram of
Fig.13. This window could be a finite size effect, but the lack of a
strong dependence with the chain lengths in the results of
Fig.14a led us to
believe that it may 
actually exist in the bulk limit. In addition, previous
studies using the strong $J_H/t$ coupling version of the FM Kondo model
have also reported  an intermediate window between phase separation 
and FM~\cite{jose}, and the analysis below
for localized spins 1/2  suggests a similar result. Studying the spin
of the ground state in this intermediate region here and in Ref.\cite{jose}
it has been observed that it 
is $finite$ i.e. apparently partial ferromagnetism appears immediately at
any finite stable density in the model with spins 3/2, at least 
working at
intermediate and large Hund couplings~\cite{comm77}. 
The transition from phase separation to FM appears to be smooth in the 
spin quantum number, which is  somewhat reminiscent of
the results in the classical limit where the tail
of the spin-spin correlation grows with continuity from small
to large as the density diminishes in the stable region.
\begin{figure}[!h]
{\rotatebox{270}{
{\resizebox{7.75cm}{!}{\includegraphics{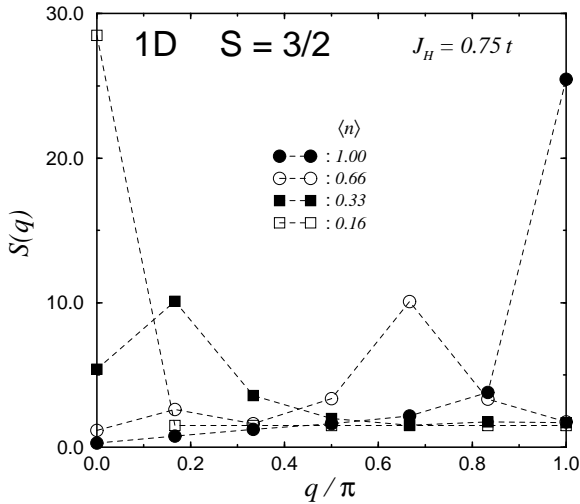}}}
}}
\caption{
$S(q)$ for the quantum FM Kondo model using localized
spins 3/2 on a chain
of 12 sites. The technique is DMRG, keeping 48 states in the
iterations. Densities and coupling are indicated.
In the spin correlations used to obtain $S(q)$, full spins 3/2 (i.e. not
normalized to 1) were used.
}
\end{figure}

Finally, let us analyze whether incommensurate correlations exist in
the spin 3/2 model as it occurs in the classical model. Fig.15 shows
$S(q)$ obtained with DMRG working on a chain of 12 sites. Although
momentum is not a good quantum number on a system with OBC,
nevertheless using the same definition of the Fourier transform of the
 spin correlation
as in the study of the classical system with periodic and antiperiodic
boundary conditions (PBC and APBC, respectively),
qualitative information about
the tendency to form incommensurate structures can be gathered.
The results of Fig.15 obtained at very small $J_H/t$ indeed show that
strong IC correlations exist in the ground state of this
model, in agreement with the results of Figs.10 and 11 for classical spins.
Then, it is concluded that the three
dominant features of the phase diagram Fig.1 (PS, FM, and IC) have an
analog in the case of the spin 3/2 quantum model. This agreement
gives support
to the believe that the results presented below in this paper for dimensions
larger than one using classical spins
should be qualitatively similar to  those corresponding to a model with
the proper quantum mechanical $t_{2g}$ degrees of freedom.

\subsection{Quantum Mechanical $t_{2g}$ S=1/2 spins}

For completeness, in this paper the special case of localized
spins 1/2 in 1D has also been studied. 
The analysis has relevance not only in the context of manganites but
also for recently synthesized one dimensional materials such as
${\rm Y_{2-x} Ca_x Ba Ni O_5}$ which have  a mobile and a localized
electron per Ni-ion. This compound has been 
studied experimentally~\cite{batlogg} and
theoretically~\cite{nio}, and upon doping interesting properties have been observed
including a metal-insulator transition.
As discussed below, the
conclusion of this subsection will be that
the results for localized
spins 1/2 are qualitatively  similar to those obtained with
classical and spin 3/2 
degrees of freedom, i.e. ferromagnetism, incommensurability, and
phase separation appear clearly in the phase diagram. 
The Hamiltonian
used for this study is as defined in Eq.(1) but now with
${\bf S}_{\bf i}$ replaced by a spin 1/2 operator (not normalized to 1). 
The technique used
to obtain ground state properties is the finite-size DMRG method on
chains with up to 40 sites, a variety of densities, and typical
truncation errors around $10^{-5}$. Some calculations were also
performed with
the Lanczos algorithm on lattices with up to 12 sites. The data
obtained with the DMRG and Lanczos methods are qualitatively similar.

The main result is contained in the phase diagram
shown in Fig.16. Three regimes
were identified~\cite{comm99}.
The region labeled FM corresponds to saturated
ferromagnetism i.e. the ground state spin is the maximum. This regime was
found using the Lanczos technique
simply searching for degeneracies between the lowest energy
states of subspaces with different total spin in the $z$-direction. 
For the IC regime, $S(q)$ was calculated
using the same definition as in subsection III.A
 but now with spin 1/2 operators instead of classical spins of length 1.
The results shown in Fig.17 suggest the presence of incommensurate
correlations at small $J_H/t$, at least at short distances.
Similarly as for the
cases of larger localized spins (Figs.10 and 15), the peak in the spin
structure factor moves smoothly from $q=\pi$ near $\langle n \rangle =1$
to $q=0$ in the FM region. The position of the peak is at $2 k_F$.
Our study also showed that 
correlated with this behavior
the charge structure factor $N(q)$ has a cusp at the same position.
It is important to clarify that in the
IC regime of Fig.17 at low temperatures, the ground state has a finite
spin but it is not fully saturated. Within the accuracy of
our study the spin varies smoothly as the  couplings and density 
are changed reaching its maximum value at the FM boundary.
\begin{figure}
{\rotatebox{0}{
{\resizebox{7.75cm}{!}{\includegraphics{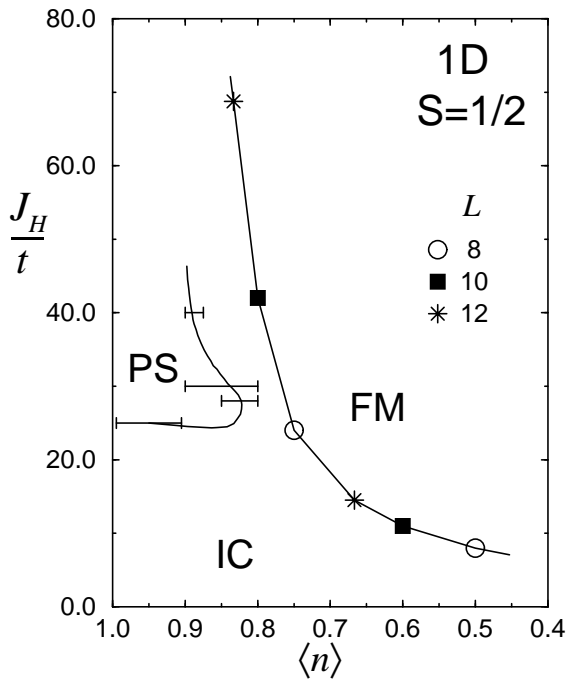}}}
}}
\caption{
Phase diagram of the FM Kondo model with $S=1/2$ localized states
obtained with the DMRG and Lanczos methods. The length of the chains is
indicated. The notation FM, PS, and IC is as in previous figures.
Note the similarity with the results of Fig.13, up to an overall scale.
}
\end{figure}
\begin{figure}
{\rotatebox{270}{
{\resizebox{7.75cm}{!}{\includegraphics{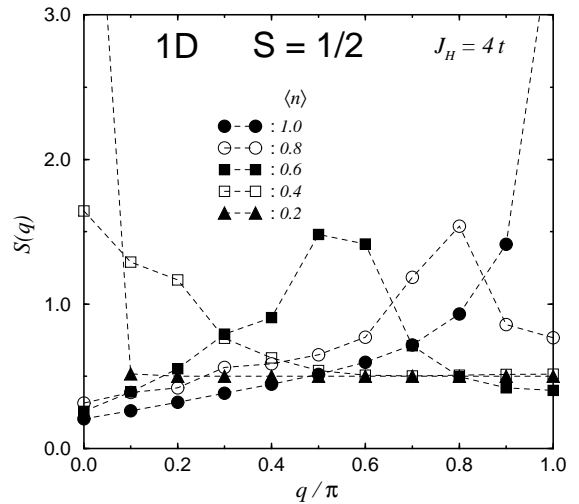}}}
}}
\caption{
$S(q)$ for the quantum FM Kondo model using localized
spins 1/2 on a chain
of 20 sites. The technique is DMRG, keeping up to 60 states in the
iterations.
Densities and coupling are indicated.
In the spin correlations used to obtain $S(q)$, full spins 1/2 (i.e. not
normalized to 1) were used.
}
\end{figure}

The tendency to phase separate in this model can be studied 
calculating $\langle n \rangle$ vs $\mu$. In the study of this section, carried
out in the canonical ensemble where $\langle n \rangle$ is fixed, 
the procedure to obtain $\mu$ involves
(i) the calculation of the ground state energy for a variety of densities,
and (ii) the addition of $-\mu {\hat N}$ to the Hamiltonian, where
${\hat N}$ is the total number operator. As $\mu$ varies, different
density subspaces become the actual global ground state. If there
are densities that cannot be stabilized for any value of $\mu$, then
such a result is compatible with phase separation in the model
(for more details see Ref.~\cite{old}).
Results are presented in Fig.18.a,b,c for a chain with
20 sites: at small $J_H/t=1$ all densities are
accessible tuning $\mu$, and the curvature of the ground state energy $E_{GS}$ vs
$\langle n \rangle$ is positive. However, working at $J_H/t = 30$ the 
density $\langle n \rangle = 0.9$ becomes unstable, and now $E_{GS}$ vs
$\langle n \rangle$ has less curvature. Fig.18d shows again density vs
chemical
potential but now on a larger chain of 40 sites. Here densities between
1 and 0.8 are unstable, in agreement with Fig.18c. A similar method to
calculate the region of phase separation is to evaluate the inverse
compressibility since a negative value for this quantity indicates phase
separation, as explained in Sec.IV.A. 
Using this procedure once again a region of
$\kappa^{-1} < 0$ was identified signaling unstable densities in the
system. From the combination of these type of analysis performed for
several
 chains, the boundaries of phase separation were estimated as shown in
 Fig.16.
Although the error bars are not negligible, the presence of phase
separation is a robust feature of the calculation and it is in 
excellent agreement with the conclusions of previous sections.
Then, irrespective of the actual value of the spin corresponding to the
$t_{2g}$ degrees of freedom the phase diagram presents universal
features,
specially robust ferromagnetism, unstable densities, and short-range
incommensurate correlations. Below in Secs.V and VI, it will be shown that this
universality can actually  be extended to include
higher dimensional clusters.
\begin{figure}
{\rotatebox{270}{
{\resizebox{5.75cm}{!}{\includegraphics{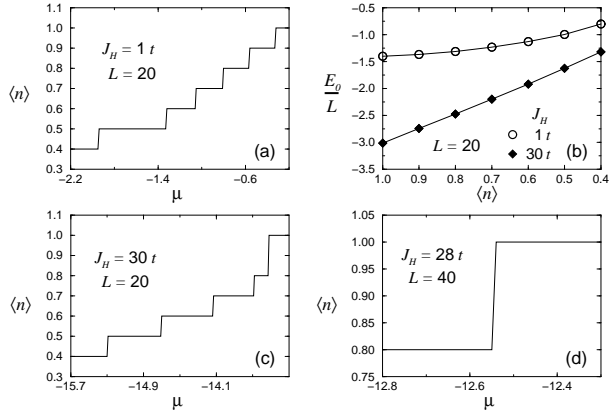}}}
}}
\caption{
Results for the FM Kondo model with $S=1/2$ localized spins using
the DMRG method.
(a) contains $\langle n \rangle$ vs $\mu$ for a chain of 20 sites at
$J_H/t=1$,
showing that all available densities are accessible;
(b) Ground state energy per site vs $\langle n \rangle$ for $J_H/t =1$
and 30. The energies for the latter were divided by 5 to make both
curves of comparable magnitude;
(c) Same as (a) but for $J_H/t=30$. Now $\langle n \rangle = 0.90$
appears unstable; (d) Results on a 40 sites chain at $J_H/t=28$. The
densities available between 1.00 and 0.80 (i.e. 0.95, 0.90, and 0.85)
 are unstable.
}
\end{figure}

\subsection{Influence of an On-site Coulomb Repulsion}

The on-site Coulombic repulsion among electrons in the conduction band
has been neglected thus far. Although it is not expected to produce
qualitative changes in the physics described before (since a large
$J_H/t$ prevents double occupancy) it would be
desirable to have some indications of its quantitative
 influence on the phase diagrams
discussed in previous sections. Unfortunately, the addition of a Hubbard repulsion $U$
complicates substantially the many-body numerical studies. 
The Monte Carlo method in
the classical limit can only proceed after the Hubbard $U$-term is
decoupled using Hubbard-Stratonovich (HS) variables. If such a procedure
is followed the simulation would run
over both angles and HS degrees of freedom, and likely a ``sign''
problem would occur.
Such a cumbersome approach will not be pursued here.
Instead the large $U/t$ limit for the case of localized spins
1/2 will be investigated using the Lanczos method, which can be applied
without major complications to the study of Hubbard-like systems. The
study will  be limited to chains due to restrictions on the size of
the clusters that can be analyzed numerically. In addition, in 
the strong coupling
limit the Hubbard model becomes the  $t-J$ model, and, thus,
the actual Hamiltonian studied here is defined as
$$
H = J \sum_i [ { {{ s}_{ i}}\cdot{{ s}_{ i+1}}} -(1/4) n_i n_{i+1} ]~~~~
$$
$$
~~~~-t \sum_{i \sigma} ( {\hat c}^\dagger_{i \sigma} {\hat c}_{i+1 \sigma} +
h.c.) - J_H \sum_i {{{\bf s}_{i}}\cdot{{\bf S}_{i}}},
\eqno(8)
$$
\noindent where ${\hat c}$ is a destruction fermionic operators which
includes a projector operator avoiding
double occupancy. Both ${\bf S}_i$ and
${\bf s}_i$ are spin-1/2 operators representing the spin of the
 localized and mobile
degrees of freedom, respectively.
The rest of the notation is standard.

Once again
studying the energy of the ground state for subspaces with different
total spin projection in the $z$-direction, the existence of 
ferromagnetism can be studied. The boundary of the region with fully
saturared ferromagnetism obtained 
using clusters with $8$ and 
$10$ sites is shown in Fig.19a. Strong ferromagnetic correlations
appear even for small values for
$J_H/t$, suggesting that as  $U/t$ grows the tendency
to favor ferromagnetism
increases, as noticed also in Ref.~\cite{jose}.
It is interesting to observe that in the region not
labeled as FM in Fig.19a a finite spin exists in the
ground state (see Fig.19b), 
in excellent agreement with results for the $U/t=0$ case
reported in Sec.IV.A and IV.B for  
 quantum mechanical $t_{2g}$ degrees of freedom (see also Ref.\cite{jose}).
In addition, a tendency towards
incommensurate correlations is observed
in this phase, also in agreement with previous results (see
Fig.19c).
Regarding the issue of
 phase separation, the results of  Fig.16 suggests that this regime should appear for
densities between $\langle n \rangle = 1.0$ and
$\langle n \rangle = 0.90$. Unfortunately these densities are not
accessible on the clusters studied in this subsection, and, thus, the confirmation of the
existence of phase separation at large $U/t$ will still require
further numerical
work~\cite{comm88}. Nevertheless, Figs.16 and 19a, which contain results with and
without the Coulomb interaction, have clear qualitative common trends.
Regarding the quantitative aspects, the Coulomb interaction changes substantially
the scales in the phase diagram
specially regarding ferromagnetism which now appears at  smaller
values of $J_H/t$~\cite{comm43}.

\begin{figure}
{\rotatebox{0}{
{\resizebox{8.5cm}{!}{\includegraphics{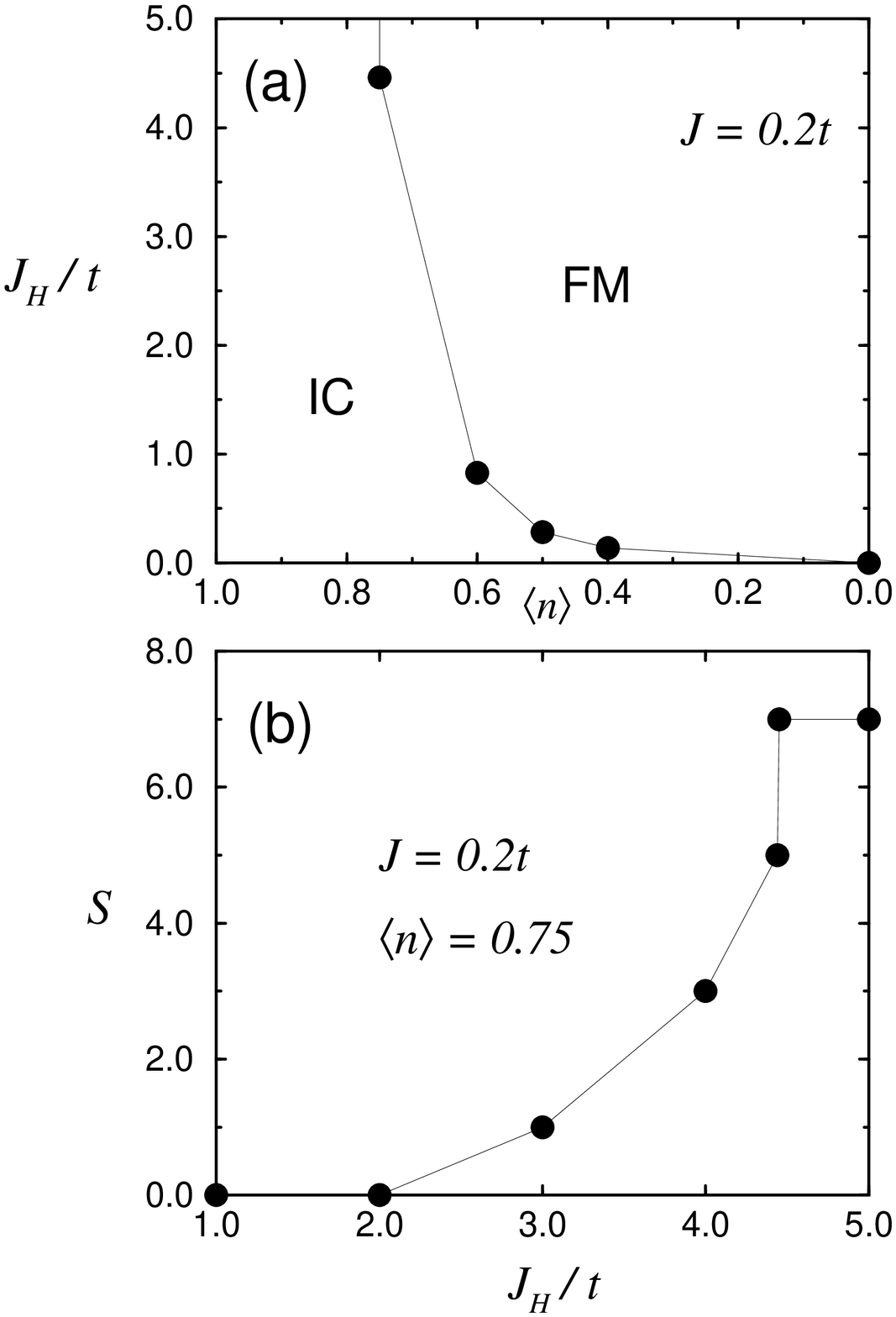}}}
}}

\vspace*{-1.4cm}
{\rotatebox{270}{
{\resizebox{6.3cm}{!}{\includegraphics{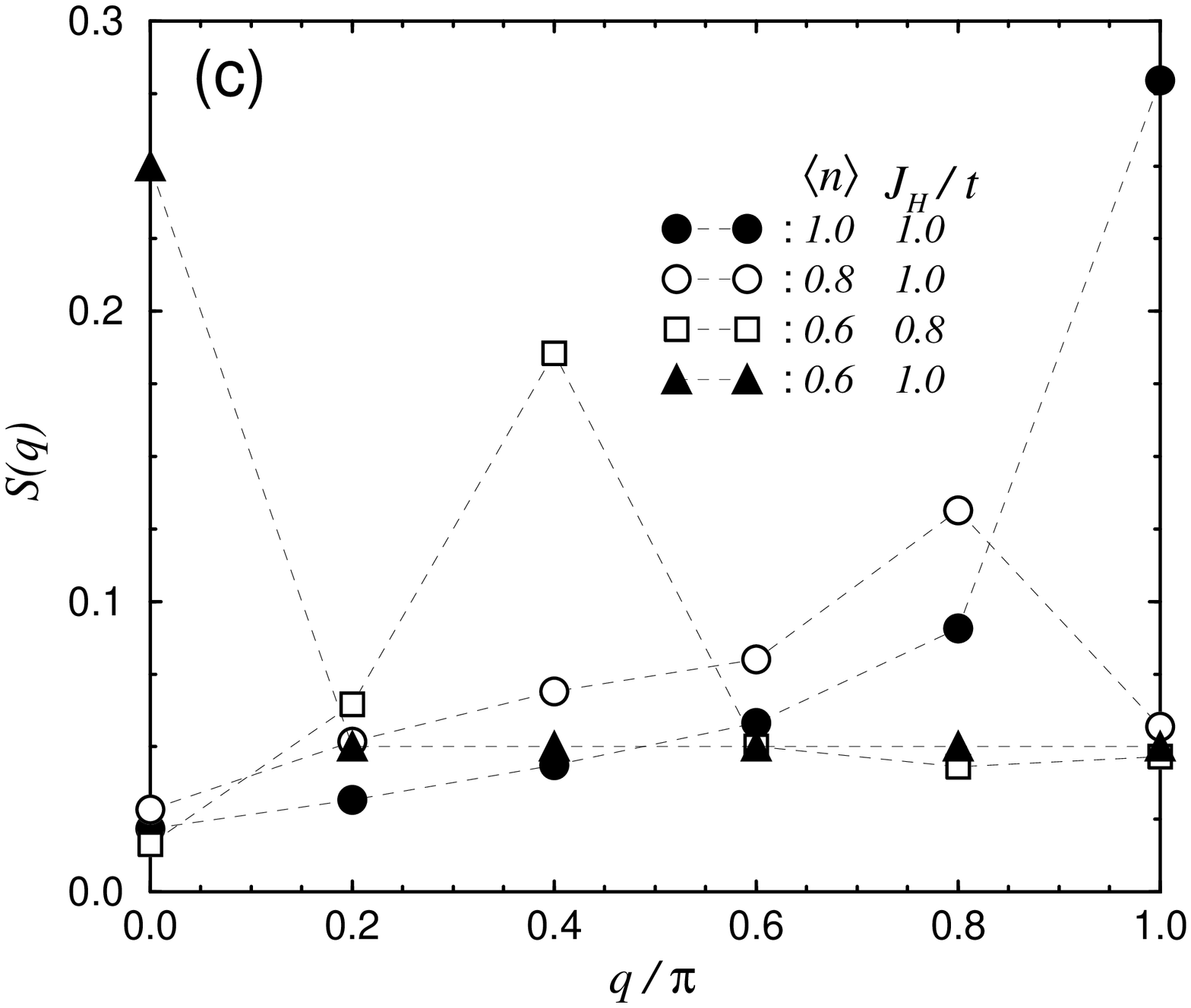}}}
}}
\caption{
Results corresponding to the FM Kondo model with
localized spins 1/2 and using the $t-J$ model for the mobile electrons
 (Eq.(8)),
obtained with the Lanczos method.
The  boundary conditions were such that a fully saturared
ferromagnetic states is stable. $J/t=0.2$ was used. (a)
shows the boundary of the FM region  using 8 and 10 sites
clusters. IC correlations  were observed at small $J_H/t$ in the
non-fully saturated ferromagnetic region; (b) the spin of the ground state
as a function of $J_H$ for the case of 2 holes on the 8 sites
cluster;
and (c) $S(q)$, the Fourier transform of the
spin-spin correlations between the localized spins,
vs momentum
 for the case of a 10 sites cluster. The densities and couplings are
indicated.
}
\end{figure}

\section{Results in D=2}

In this section the computational results that led us to propose Fig.2 as the
phase diagram of the FM Kondo model in two dimensions are presented.
Results in 2D are particularly
important since manganite compounds with layered structure
have been recently synthesized~\cite{moritomo}.
\subsection{Ferromagnetism}

The search for ferromagnetism in the ground state of the 2D clusters
was carried out similarly as in 1D for the case of
classical $t_{2g}$ spins. The real space
spin-spin correlations (between
those classical spins) was monitored, as well as its Fourier transform
$S(q)$ at zero momentum. Couplings $J_H/t = 1,2,4,8,$ and $16$ were
particularly analyzed. Typical results are presented in Fig.20 where
the spin correlations are presented at a fixed coupling, parametric with
the electronic density. It is clear that the tails of the correlations
are very robust, and the ferromagnetic correlation length exceeds the
maximum distance, $d_{max}$, available on the $6 \times 6 $ cluster. 
Plotting the spin correlation at $d_{max}$ vs $\langle n \rangle$, 
an estimate of the critical density for ferromagnetism can be obtained.
Combining results from a variety of clusters and boundary conditions,
the FM boundary of Fig.2 was constructed. Note that for this analysis the
use of open boundary conditions seem to be the optimal i.e. using
other boundary conditions kept the ferromagnetic character of the 
system at short distances but modify the correlations at large distances,
as it occurs for non-closed-shell BC in 1D.
\begin{figure}[h]
{\rotatebox{0}{
{\resizebox{7.75cm}{!}{\includegraphics{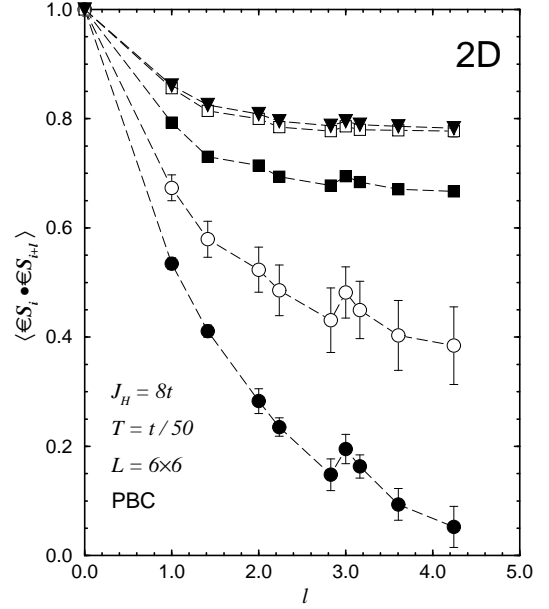}}}
}}
\vspace*{-1cm}
\caption{
Spin-spin correlations (among the classical spins) vs. distance
obtained at $J_H/t=8$, $T=t/50$, and
using periodic boundary conditions on a $6 \times 6$ cluster.
Full circles, open circles, full squares, open squares, and full
triangles correspond to densities
$\langle n \rangle =$ 0.807, 0.776, 0.750, 0.639, and
0.285, respectively.
}
\end{figure}

The influence of the lattice size can be estimated by studying
$S(q)$ vs temperature for several clusters. Representative results
are shown in Fig.21. They were obtained at $J_H/t=16$ and a density
close to the PS-FM 
boundary. At low temperature $S(q=0)$ clearly
grows with the lattice size due to the strong ferromagnetic
correlations. However, as in the case of 1D, the Mermin-Wagner theorem
forbids long-range ferromagnetism in 2D at finite temperature and, thus,
$S(q)$ should converge to a finite constant
if the lattice sizes are further increased beyond those currently accessible,
working at a fixed
 finite temperature. The verification of this subtle detail is
beyond the scope of this paper, and should not confuse the readers: the
presence of very strong ferromagnetic correlations at low temperature in
the FM Kondo model is clear in the present numerical study and it is
likely that
small couplings in the direction perpendicular to the planes
 would stabilize ferromagnetic order at
a finite temperature.
\begin{figure}
{\rotatebox{0}{
{\resizebox{7.75cm}{!}{\includegraphics{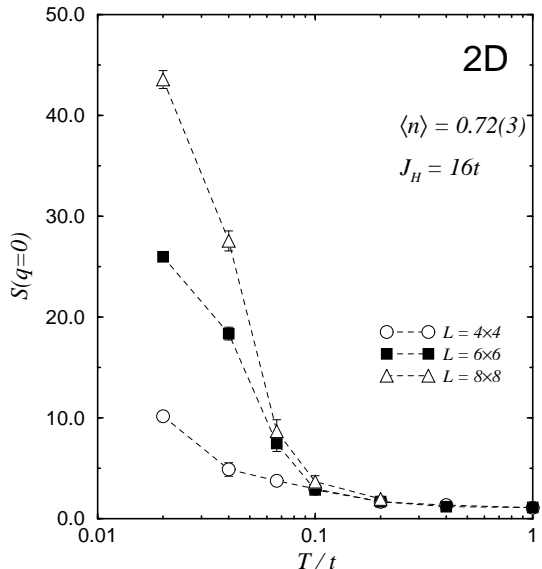}}}
}}
\vspace*{-2cm}
\caption{
$S(q)$ at zero momentum vs. temperature, working at
 $\langle n \rangle = 0.72 \pm 0.03$, $J_H/t=16$, 
using open boundary conditions,
and for the two-dimensional cluster sizes shown.
}
\end{figure}

\subsection{Phase Separation}

The computational analysis carried out in this subsection show
that the phenomenon of phase separation occurs not only in one dimension
but also in
two dimensions (and higher). This result indicates that the 
unstable densities found
in Secs.III and IV are not a pathology of 1D clusters but its existence is
generic of the FM Kondo model. 

Typical numerical results in 2D clusters at low temperature
 are shown in Fig.22a at $J_H=8t$. Data from a variety of 
cluster sizes and boundary conditions are presented.
Although the results corresponding to the lowest density 
in the discontinuity in $\langle n \rangle$ 
are somewhat
scattered due to size effects, the existence of such discontinuity
is clear  from the figure. Fig.22b contains 
similar results but now for $J_H/t=4$.
Once again, $\langle n \rangle$ is discontinuous signaling the
presence of phase separation in the low temperature regime of
the FM Kondo model. Calculating the discontinuity in $\langle n \rangle$ for
several couplings 
the boundary of the phase separated
 regime in the 2D phase diagram of Fig.2
was established. Note that the scales in computer
time needed to achieve convergence are very large near the critical
chemical potential where frequent tunneling events between the two
minima slow down the simulations. For $J_H/t < 4$, it becomes
difficult to distinguish between an actual discontinuity in $\langle n
\rangle$ and a very rapid crossover and, thus, in the 2D
phase diagram (Fig.2) the boundary of phase separation
 at small Hund coupling is 
not sharply defined.
\begin{figure}
{\rotatebox{270}{
{\resizebox{6.5cm}{!}{\includegraphics{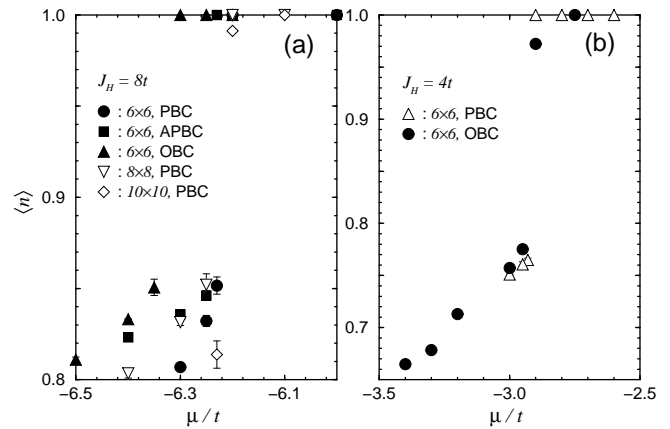}}}
}}
\caption{
$\langle n \rangle$ vs $\mu$ on two-dimensional clusters
and at temperature $T=t/50$, to illustrate the presence of phase separation
in the 2D ferromagnetic Kondo model with classical spins. (a) contains results at
 $J_H/t = 8$
and a variety of clusters and boundary conditions. (b) corresponds
to $J_H/t=4$,  a $6 \times 6$ cluster, and using both periodic
and open boundary conditions.
}
\end{figure}

\subsection{Incommensurate Correlations}

The regime of small coupling $J_H/t$ is not ferromagnetic,
according to the behavior of $S(q)$ at zero momentum, and does not
correspond to phase separation since all densities  are stable.
It may occur that robust incommensurate spin
correlations exist here, as it occurs in 1D. 
$S(q)$ is presented in Fig.23 for two representative
couplings and a large range
of densities. The antiferromagnetic peak at $(\pi,\pi)$
is rapidly suppressed as $\langle n \rangle$ decreases,
 and the position of the maximum in $S(q)$ moves along the
$(\pi,\pi)$--$(\pi,0)$ line (and also along $(\pi,\pi)$--$(0,\pi)$ by
symmetry). The intensity of the peak is rapidly reduced as 
it moves away from $(\pi,\pi)$, and the IC character of the correlations
\begin{figure}[h]
{\rotatebox{0}{
{\resizebox{7.75cm}{!}{\includegraphics{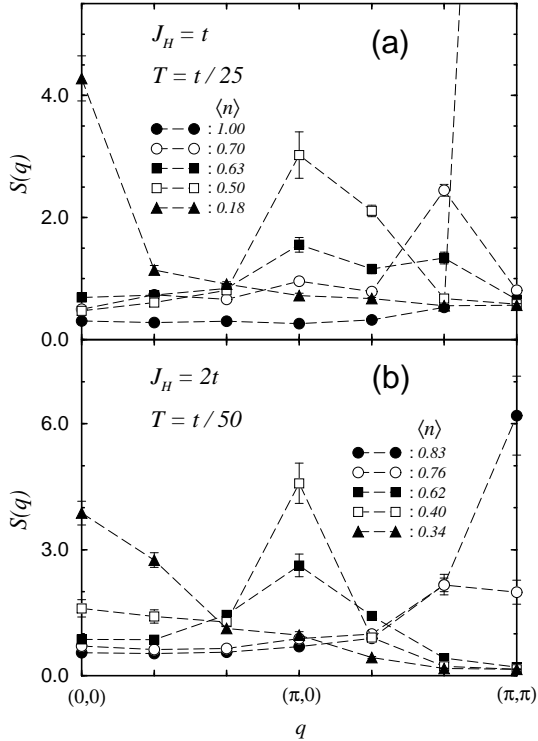}}}
}}
\caption{
$S(q)$ vs momentum on a $6 \times 6$ cluster and
at (a) $J_H/t=1$ and $T=t/25$, and (b) $J_H/t=2$ and $T=t/50$.
In both cases open boundary conditions were used. The densities
are indicated.
}
\end{figure}
is apparently short-range. However, the numerical
study of IC phases are notoriously
affected by lattice sizes, and thus at this stage it can only be
claimed that a tendency to form IC spin patterns has been detected
in 2D clusters, without a firm statement regarding their short- vs.
long-range character. As $\langle n \rangle $ is further reduced, the
peak position
reaches $(\pi,0)$ and $(0,\pi)$ with substantial intensity. Reducing further
the density, a rapid change into the ferromagnetic phase was observed. 
At a larger $J_H/t$ coupling such as 3, a more complicated
IC pattern was detected with regions where $S(q)$ peaked simultaneously
at $(\pi,0)$-$(0,\pi)$ and $(0,0)$. Further work is needed to clarify
the fine details of the spin arrangements in this regime, but
nevertheless the results given here are enough to support the claim
that a tendency to form incommensurate correlations exists in the
ground state of the 2D FM Kondo model~\cite{hamada}.

\section{Results in D=$\infty$}

The existence of phase separation and ferromagnetism in the
ground state of the FM Kondo model can also be studied in
the limit of $D= \infty$. 
The Dynamical Mean Field equation~\cite{furukawa} 
is solved iteratively
starting from a random spin configuration,
and as a function of temperature and density three solutions have
been observed having AF, FM, and paramagnetic character. 
Efforts
were concentrated on a particular large coupling $J_H/W=4.0$
studying the temperature dependence of the results, where 
$W$ is the half-width of the semicircular density of states 
$D( \epsilon ) = (2 / \pi W) \sqrt{ 1 - (\epsilon / W)^2}$
for the  $e_g$ electrons.
Partial results
are contained in Fig.24a. The presence of ferromagnetism at finite
doping and antiferromagnetism at half-filling are
quite clear in the calculations. Close to half-filling and at
low temperature, the density $\langle n \rangle$ as a function
of $\mu$ was found to be discontinuous, in excellent agreement with
the results already reported in $D=1$ and 2. Fig.24b provides a typical example
obtained at $T/W = 0.0003$ (results at $T/W = 0.002$ were already
presented in Ref.~\cite{previous}). The phase
separation in this figure is between 
antiferromagnetic and ferromagnetic regions. However, in
 Fig.24b note that at a slightly
larger temperature the separation occurs between hole-poor antiferromagnetic
and hole-rich $paramagnetic$ regions.

For completeness, in Fig. 25 the density of states $A(\omega)$ for
the AF and FM phases is shown at $J_H/W=2$ and $T/W=0.005$
(for details of the calculation see Ref.\cite{furukawa}). The
critical chemical potential where the AF and FM phases
coexist is $\mu_c \sim -1.40 W$.
$A(\omega)$ in Fig.25 is calculated at $\mu=\mu_c$ for both phases.
In the two cases the density of states splits into
upper and lower bands due to the large Hund coupling. 
The width of the upper and lower bands is wider for the FM phase,
which causes a narrower gapped region centered at $\omega \sim 0$.
Let us now consider the process of hole doping starting at
 $\langle n \rangle = 1$ and decreasing $\mu$.
In the AF  phase at $\mu \ge \mu_c$, the chemical potential
lies in the gap.
However,  at $\mu \le \mu_c$ the chemical potential
is located already inside the lower band of the FM phase,
since this band is wider than in the AF phase.
This suggest that 
before the lightly doped AF phase is realized in the system 
by decreasing $\mu$, the
FM phase is instead stabilized.
Thus, the discontinuous change from the AF to FM phases at $\mu=\mu_c$
also causes a jump in the carrier number. 
This discontinuity occurs
only when the bandwidth of the AF phase is considerably narrower
than that of the FM phase. For this reason phase
separation exists only  in the large $J_H$ region.

Summarizing, at very low temperature
there is a remarkable
agreement between the $D=\infty$ and $D=1,2$ results which were obtained
using substantially
different numerical techniques. Such an agreement give us
confidence
that the phase separation effect discussed here is not pathological
of low dimensions or induced by approximate algorithms but it is
intrinsic of the physics of the FM Kondo model, and likely it exists
in dimension $D=3$ as well.

\begin{figure}[h]
{\rotatebox{270}{
{\resizebox{6.75cm}{!}{\includegraphics{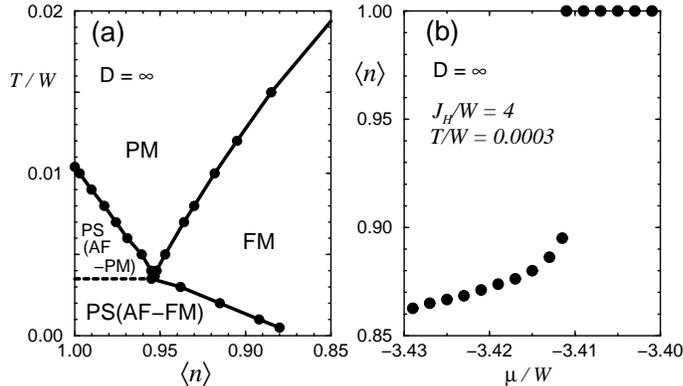}}}
}}
\caption{
(a) Phase diagram in the $D=\infty$ limit working at
$J_H/W = 4.0$. The ``PS(AF-PM)'' region denotes phase separation
(PS) between a hole-poor antiferromagnetic (AF) region, and
a hole-rich paramagnetic (PM) region. The rest of the notation
is standard; (b) Density $\langle n \rangle$ vs $\mu/W$ obtained
in the $D=\infty$ limit, $J_H/W=4.0$, and $T/W=0.0003$. The
discontinuity in the density is clear.
}
\end{figure}
%
\begin{figure}[!h]
{\resizebox{8.0cm}{!}{\includegraphics{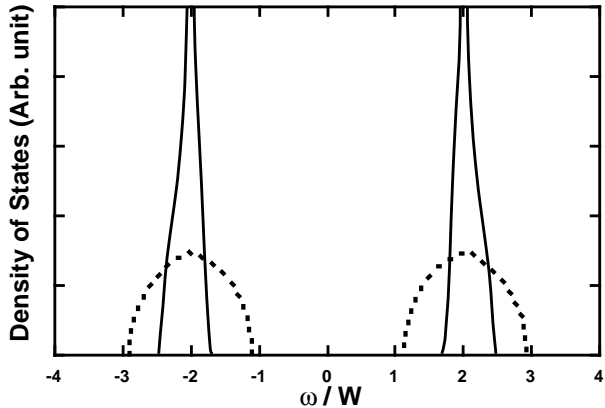}}}
\caption{
Density of states in the $D=\infty$ limit corresponding to the
antiferromagnetic (solid line) and ferromagnetic (dotted line) solutions,
working at $J_H/W = 2.0$ and $T/W = 0.005$. The chemical potential
is tuned to be at its critical value $\mu_c \sim -1.40 W$.
}
\end{figure}
\section{Results in D=3}

Results in three dimensions for large enough clusters are difficult to obtain 
with the Monte Carlo algorithm used in this paper.
The reason is that the CPU time 
needed to diagonalize exactly the problem of an electron
moving in a fixed spin background grows rapidly with
the number of sites. Nevertheless, 
studies using $4^3$ clusters carried out  as part of this project
 have shown clear
indications of strong ferromagnetic correlations in a region of parameter
space compatible with those found in one and two dimensions where FM dominates, and thus
it is reasonable to assume that there are ferromagnetic phases in the
FM Kondo model in all dimensions from 1 to $\infty$. Regarding phase
separation, the studies on $4^3$
clusters cannot provide conclusive evidence due to the presence of
intrinsic gaps in $\langle n \rangle$ vs $\mu$ caused by
size effects. But once again by simple continuity between $D=1,2$ and
$D=\infty$, the existence of phase separation in $D=3$ is strongly
suggested by our results.

In spite of the size limitations of studies in three dimensions, it is
possible to obtain useful information about the actual value of
the critical temperature in the limit of $J_H = \infty$ i.e.
working in a
region which corresponds to a fully saturated ferromagnetic state at
zero temperature. In this limit
the problem is  simplified since for the mobile electrons only the
spin component in the direction of the classical spin survives, as
discussed in Sec.II where the
effective model (``complex double exchange'') was described.
The absence of a spin index reduces substantially the CPU time for
diagonalization in the numerical algorithm, and allowed us to study clusters
with $6^3$ sites using the Hamiltonian Eq.(3). 
Measuring the spin-spin
correlation (among the classical spins) in real space 
and reducing the temperature,
it is possible to study at what temperature, $T^*$,
 such correlation becomes nonzero  at the maximum
distance available in a  $6^3$ cluster.
For temperatures higher than $T^*$ the spin correlations can be
accommodated inside the cluster and, thus, the ferromagnetic correlation
length, $\xi^{FM}$, is $finite$. Using this idea, upper bounds 
on the critical
temperature can be obtained using $T^* > T_c^{FM}$. In addition,
it is reasonable to assume that in a 3D system the growth of $\xi^{FM}$
with $T$ is very rapid once it starts,
and  $T^*$ itself may actually provide a good
estimate of the critical temperature in the bulk.
\begin{figure}
{\rotatebox{0}{
{\resizebox{7.75cm}{!}{\includegraphics{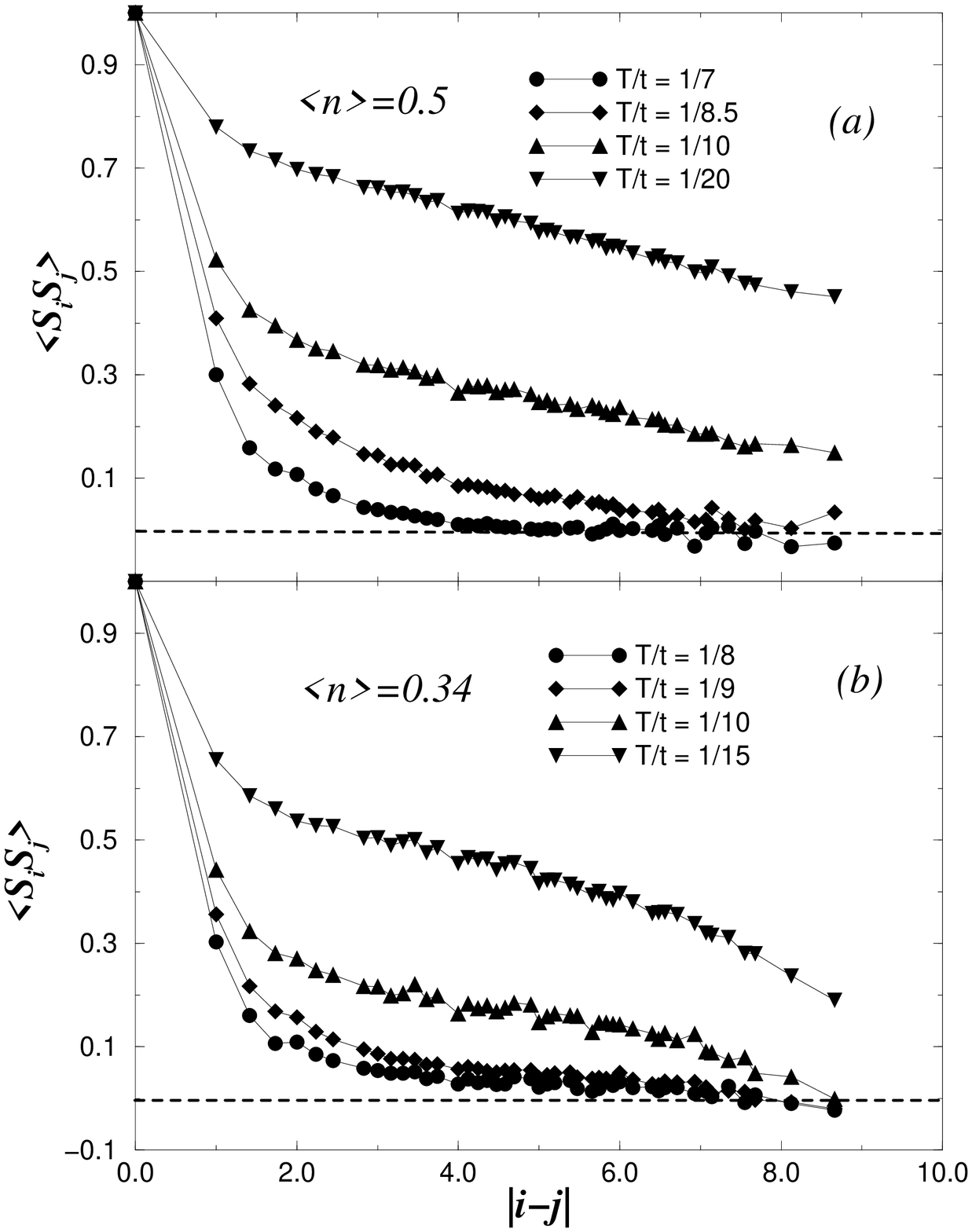}}}
}}
{\rotatebox{0}{
{\resizebox{7.75cm}{!}{\includegraphics{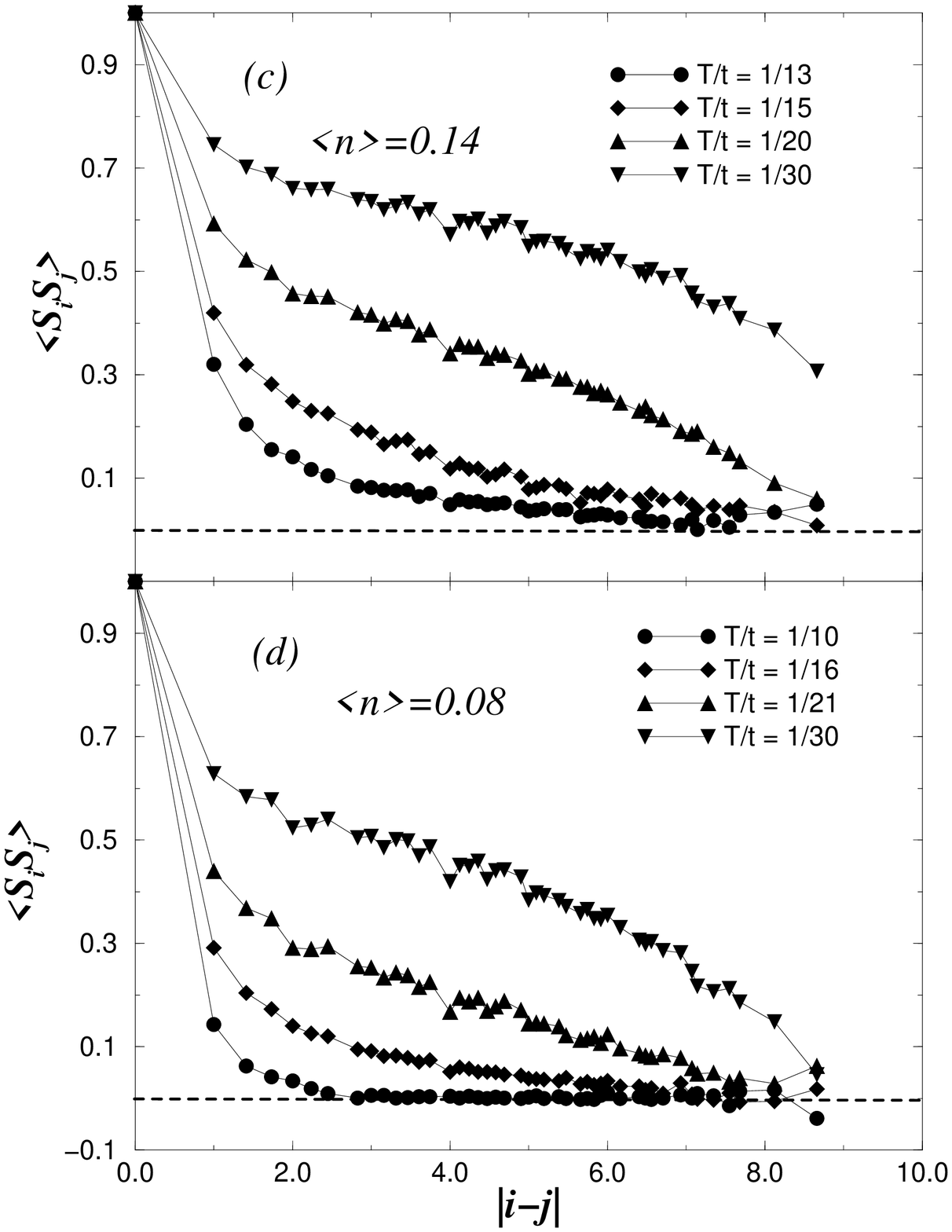}}}
}}
\caption{
Spin-spin correlations (among the classical spins) vs distance for
several temperatures (as indicated) working on a $6^3$ cluster and using
the Monte Carlo technique. (a) contains
results for $\langle n \rangle = 0.50$, (b) for $\langle n \rangle
= 0.34$, (c) for $\langle n \rangle = 0.14$, and (d) for
$\langle n \rangle = 0.08$.
}
\end{figure}

Figs.26.a-d contain the spin-spin correlations at $J_H = \infty$ on
the $6^3$ cluster parametric with temperature. Results for four
densities are given, and only some representative temperatures are
shown. The spin correlations on a $4^3$ cluster (not shown)
are in good agreement with those provided in Fig.26.
Based on this information the temperature where the ferromagnetic
correlations reach the boundary with a nonzero value
can be obtained with reasonably small error bars. The results are shown 
in Fig.27a. Once again, assuming a rapid increase of $\xi^{FM}$ as
the temperature is reduced, the results of Fig.27a can be considered as 
a rough estimate of the actual critical
temperature. As anticipated in Sec.III.A, $T_c^{FM}$ is maximized
at $\langle n \rangle = 0.50$. In Fig.27b, $S(q)$ at zero momentum is 
presented as a function of temperature for the four densities
used in Fig.26. A rapid growth is observed at 
particular temperatures which are  
compatible with those obtained in Fig.27a using the
tail of the real space correlations.
The present results are qualitatively similar to
those obtained using high temperature expansions (HTE)~\cite{high}, 
although our
estimates for the critical temperatures
are  smaller. For instance, at $\langle n \rangle =
0.50$ the HTE prediction gives $T_c^{FM} \sim 0.16t$, about a factor 1.5
larger than our result. 
\begin{figure}[h]
{\rotatebox{0}{
{\resizebox{8.5cm}{!}{\includegraphics{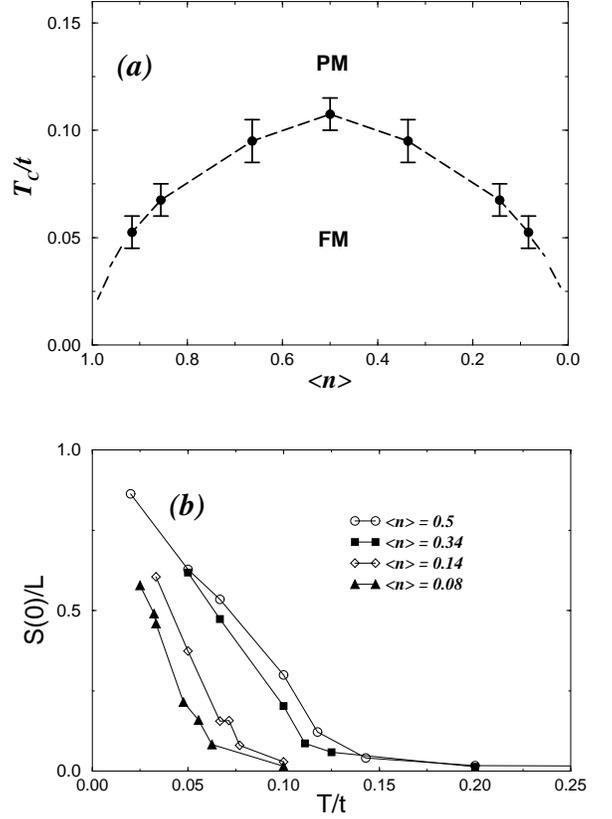}}}
}}
\caption{
(a) Upper bound for the critical temperature vs density deduced from
spin correlations on $6^3$ clusters, working at $J_H = \infty$; (b)
$S(q=0)/L$ vs temperature at several densities, obtained with the Monte
Carlo method on a $6^3$ cluster at $J_H/t = \infty$.
}
\end{figure}

To obtain the critical temperature in degrees, an estimate of
$t$ is needed. Results in the experimental literature
 for the $e_g$ electrons bandwidth range
from $BW \sim 1~eV$~\cite{bandwidth} to $BW \sim 4 eV$~\cite{sarma}. 
Assuming a dispersion $\epsilon_{\bf p} = -2t(cosp_x + cosp_y +
cosp_z)$, the hopping amplitude should be
$t = BW/12$ i.e. between $0.08$ and $0.33~eV$. With this result 
our estimate for the critical temperature is roughly  between
$T_c^{FM} \sim 100~K$ and $400~K$, which are
within the range observed
experimentally. Then, in our opinion it is possible to
obtain realistic values for
$T_c^{FM}$ using purely electronic
models, in agreement with 
other calculations~\cite{furukawa,high}.

\section{Conclusions}

In this paper the phase diagram of the ferromagnetic Kondo model for
manganites was investigated. Using a wide variety of computational
techniques that include Monte Carlo simulations, Lanczos and DMRG
methods, and the Dynamical Mean Field approach,
regions in parameter space with (i) robust ferromagnetic
correlations, and (ii) phase separation between hole-poor
antiferromagnetic and hole-rich ferromagnetic regions were identified.
In addition,  incommensurate spin correlations were observed
in dimension 1 and 2 at small $J_H/t$.
The critical temperature towards ferromagnetism for the case of 
a three dimensional lattice was also estimated, and the results  are
compatible with experiments for manganites. 
The agreement between the results obtained with different computational techniques,
working at several spatial dimensions, and using both classical
and quantum mechanical localized spins in the case of chains lead us to
believe that the conclusions of this paper are robust and they represent
the actual physics of the ferromagnetic Kondo model.

The novel regime of phase separation is particularly interesting, and possible 
consequences of its existence in manganites can be envisioned.
Experimentally, phase separation  can be detected using neutron diffraction
techniques if the two coexisting 
phases have different lattice parameters as it occurs in
${\rm La_2 Cu O_{4 + \delta}}$,
a ${\rm Cu}$-oxide 
with hole-rich and hole-poor regions~\cite{rada}. NMR and NQR
spectra, as well as magnetic susceptibility measurements,
can also be used to detect phase separation~\cite{hammel,bao}
since a splitting of the
signal appears when there are two different environments for the ions.
Note also that in the regime where AF and FM coexist $S({ q})$ presents a
two peak structure, one located at the AF position and the other at zero
momentum. This also occurs in a canted ferromagnetic state and,
thus, care
must be taken in the analysis of the experimental data.
Actually, recent experimental results by Kawano et al.~\cite{kawano} are in
qualitative agreement with the results of
Fig.24a since these authors observed a
reentrant structural phase transition accompanied by 
``canted ferromagnetism'' below $T_c^{FM}$, at $0.10 < x < 0.17$
in ${\rm La_{1-x} Sr_x Mn O_3}$.
In addition the polaron-ordered
phase  reported by Yamada et al.~\cite{yamada} can be reanalyzed using
the results of this paper
since it is known that the AF phase in 3D manganites
is orthorhombic while the FM
is pseudo-cubic. The formation of a lattice superstructure may stabilize the
magnetic tendency to phase separate and minimize lattice distortions.

Note, however, that phase separation may manifest itself 
as in ``frustrated phase separation'' scenarios~\cite{tj3}:
Since
the Coulombic interaction between holes was not explicitly included it
is possible that in realistic situations phase separation may be
replaced by the formation of complex structures, such as the stripes
observed in cuprates~\cite{tj2,rice,tranquada}. Thus, it is reasonable
to speculate that these stripes could also appear in the insulating
regime of the manganites and they should be detectable using
 neutron scattering techniques.

On the theoretical side, future work will be directed
to the analysis of
the influence of phonons and orbital degeneracy into the phase diagram
observed in the present paper, specially regarding phase separation, 
as well as  the calculation
of dynamical properties for the models investigated here. 
Work is in progress along these fronts to complete a qualitative
understanding of the phase diagram corresponding to models for the
manganites beyond the double exchange model.

\section{Acknowledgments}

We thank K. Hallberg, J. Riera and S. Kivelson for useful conversations.
E. D. and A. M. are supported by the 
NSF grant DMR-9520776. 
S. Y. is supported by the Japanese Society for the Promotion of Science.
J. H. is supported by the Florida State grant E\&G 502401002. 
A. L. M. acknowledges the financial support
of the Conselho Nacional de Desenvolvimento Cient\'\i fico
e Tecnol\'ogico (CNPq-Brazil)

%

\begin{thebibliography}{99}
\bibitem{comm77}
Actually working at very small
$J_H/t$ and close to half-filling  a ground state singlet was
stabilized but it is unclear to what extend this result survives the
bulk limit.
\bibitem{comm99} The $S=1$ chain has a nonzero Haldane gap at
half-filling. However, a low density of holes closes the gap
as recently observed numerically~\cite{nio}. Then, a ``spin-gap'' regime is
not included in the phase diagram of the model with
localized $S=1/2$ spins.
\bibitem{comm88} Nevertheless it was observed that the ground state energy vs
density is almost linear in the range corresponding to 0, 2 and 4 holes,
and thus phase separation may appear if larger clusters could be studied.
\bibitem{comm43} Note that a trivial factor 2 in the overall $J_H$
scale can be accounted for by considering that the mobile spin in Eq.(8)
is 1/2, while in Sec.IV.B for the conduction electrons no factor 1/2
was used in the Hund term.
\end{thebibliography}
\end{document}